\documentclass[aps,preprint,nofootinbib,showpacs,showkeys,10pt]{revtex4-2}
\usepackage[utf8]{inputenc}
\usepackage[english]{babel}
\usepackage{natbib}
\usepackage{dirtytalk}
\usepackage{amsmath,amsthm}            
\usepackage{extarrows}
\usepackage{amssymb}
\usepackage{nicefrac}
\usepackage{cancel}
\usepackage{fancyref}
\usepackage{hyperref}
\usepackage{siunitx} 
\usepackage{graphicx}                   
\usepackage{subfigure}
\usepackage{float} 
\restylefloat{figure}
\restylefloat{table}
\usepackage{xcolor}
\usepackage{braket} 
\DeclareMathOperator*{\SumInt}{%
\mathchoice%
  {\ooalign{$\displaystyle\sum$\cr\hidewidth$\displaystyle\int$\hidewidth\cr}}
  {\ooalign{\raisebox{.14\height}{\scalebox{.7}{$\textstyle\sum$}}\cr\hidewidth$\textstyle\int$\hidewidth\cr}}
  {\ooalign{\raisebox{.2\height}{\scalebox{.6}{$\scriptstyle\sum$}}\cr$\scriptstyle\int$\cr}}
  {\ooalign{\raisebox{.2\height}{\scalebox{.6}{$\scriptstyle\sum$}}\cr$\scriptstyle\int$\cr}}
}
\usepackage[compat=1.1.0]{tikz-feynman}
\usepackage{mathrsfs}
\usepackage{slashed}

\newcommand{\be}{\begin{equation}}
\newcommand{\ee}{\end{equation}}
\newcommand{\bea}{\begin{eqnarray}}
\newcommand{\eea}{\end{eqnarray}}

\begin{document}

\title{On the Propagator of a Massive Charged Vector Boson in a Magnetic Field: Ritus Eigenfunction Method}

\author{Manuel Emiliano Monreal Cancino}

\email{emilianomonreal@ciencias.unam.mx}

\author{Angel Sánchez}

\email{ansac@ciencias.unam.mx}

\affiliation{Facultad de Ciencias, Universidad Nacional Aut\' onoma de M\' exico, Apartado Postal 50-542, Ciudad de M\'exico 04510, M\'exico.}


\begin{abstract}
In this work, we derive the propagator of a massive charged vector boson in the presence of a homogeneous and constant magnetic field of arbitrary strength, working in the unitary gauge and in the mostly minus metric. The propagator is constructed using the Ritus eigenfunction method, which allows for an explicit treatment of Landau-level quantization and spin degrees of freedom. We present a detailed analysis of the polarization vectors for all Landau levels. Using the Ritus representation, we formulate and derive the LSZ reduction formula for massive charged vector bosons in a magnetic field background, providing a useful tool for the calculation of self-energies and radiative corrections with external charged states. Furthermore, we establish a systematic connection between the Ritus' eigenfunctions and Schwinger proper-time representations of the propagator, identifying Schwinger's phase and exhibiting a slight discrepancy with previous results in the literature that arise in the unitary gauge, highlighting the importance of a careful treatment of spin and gauge structures in an external magnetic field.
\end{abstract}

\maketitle
\section{Introduction}
\label{sec.intro}

The behavior of charged particles in the presence of strong magnetic fields has become a subject of great interest across several areas of high-energy, nuclear, and astrophysical physics. Strong electromagnetic fields naturally arise in a variety of physical environments, including non-central heavy-ion collisions~\cite{ASystematicStudy_zhong, EstimateOfTheMagneticField_skokov, TheEffectsOfTopological_kharzeev}, compact astrophysical objects such as magnetars~\cite{MagneticThermal_pons}, and the early Universe~\cite{MagneticFields_grasso}. In these physical scenarios, external magnetic fields can significantly alter the structure of quantum states and the dynamics of fundamental interactions. Some examples include  modifications of the superconducting phases in Quantum Chromodynamics (QCD)~\cite{SuperconductivityOfQCD_chernodub, StatisticalPhysics_landau}, the magnetic catalysis of chiral symmetry breaking~\cite{Shovkovy:2012zn, Ferrer:2009nq}, and charge separation phenomena such as the chiral magnetic effect~\cite{TopologyMagneticField_kharzeev, StronglyInteractingMatter_adhikari}.

Alongside the theoretical developments, there is increasing experimental interest in probing quantum systems under extreme electromagnetic field strengths. High-intensity laser facilities and related experimental programs aim to test fundamental predictions of quantum electrodynamics (QED) in regimes where nonlinear and nonperturbative effects become relevant~\cite{LetterOfIntent_abramowicz, OnSeminalHEDP_meuren}. These efforts provide a unique opportunity to investigate the structure of the QED vacuum under extreme conditions and motivate the development of reliable theoretical tools applicable to a wide range of field strengths~\cite{ExtremelyHighIntensity_dipiazza, ElectronMassShift_dipiazza}.

A useful theoretical ingredient in the description of quantum processes in external electromagnetic fields is the propagator of charged particles~\cite{ElectroweakProcesses_kuznetsov, ChargedMassive_iablokov}. In particular, the propagator of massive charged vector bosons in a homogeneous and constant magnetic field plays a crucial role in the analysis of processes involving neutrinos and electroweak interactions~\cite{NeutrinoDecay_kuznetsov}. Previous studies have employed different techniques to evaluate such propagators and analyze their phenomenological consequences. Early applications based on the Ritus eigenfunction method focused on neutrino propagation and refraction in magnetized media~\cite{NeutrinoSelfEnergy_elizalde, NeutrinoPropagation_elizalde}, typically within the Lowest Landau Level (LLL) approximation. In Ref.~\cite{NeutrinoDispersion_kuznetsov}, the authors demonstrated that higher Landau levels can provide non-negligible contributions, invalidating the dominance of the LLL in several physical processes. Similar works analyze neutrino decay and absorption processes in strong magnetic fields, sometimes yielding different results~\cite{HighEnergyNeutrino_erdas, NeutrinoAbsorption_sahu}, while later analyzes clarified aspects of the neutrino self-energy corrections~\cite{Neutrinoselfenergy_erdas, HighEnergyNeutrino_kuznetsov}. In most of these studies, calculations were performed in the Feynman gauge, where the tensor structure of the propagator simplifies considerably. In other works, ghost-field contributions to self-energy corrections were not explicitly discussed~\cite{HighEnergyNeutrino_erdas, MagneticField_erdas, NeutrinoDecay_kuznetsov}, while other analyses demonstrated their relevance even within the Feynman gauge~\cite{HighEnergyNeutrino_kuznetsov, NeutrinoSelfEnergyRxi_erdas}. Moreover, gauge-dependent tensorial structures can be phenomenologically important, for example in precision determinations of the decay-width difference between neutral and charged $\rho$ mesons, which are required to match the experimental accuracy of the muon anomalous magnetic moment~\cite{Measurement_aguillard, WidthDifference_flores-baez, BeyondScalar_flores-baes}. These considerations call for a systematic study across different gauge choices and, in particular, an explicit analysis beyond the Feynman gauge, including the unitary gauge limit.

\newpage

The propagator of a massive charged vector boson in a constant and homogeneous magnetic background has been derived using a variety of formalisms, including the Fock–Schwinger proper-time method~\cite{MagneticField_erdas}, canonical quantization approaches~\cite{VectorBoson_nikishov}, extensions of Ritus's eigenfunction technique to spin-1 particles~\cite{NeutrinoSelfEnergy_elizalde}, and modified proper-time formulations~\cite{ChargedMassive_iablokov}. Although these methods are formally equivalent, they differ in their computational convenience and suitability for some calculations. In particular, representations involving non-diagonal spin structures or implicit momentum-space formulations can obscure the physical interpretation of polarization states and complicate perturbative calculations~\cite{ChargedMeson_scoccola}. The availability of multiple representations of charged-particle propagators is therefore essential, as different physical scenarios may favor different formulations. In the literature, derivations of the charged vector boson propagator based on the Ritus method are typically performed using the mostly-positive metric or rely on non-diagonal spin bases~\cite{NeutrinoSelfEnergy_elizalde, ChargedMeson_scoccola}. These features could make the extraction of physical observables and the implementation of standard field-theoretical tools difficult. In contrast, the approach adopted in this work employs a diagonal spin basis and is formulated in the mostly-minus metric.

Finally, the calculation of radiative corrections in a magnetic background requires a consistent prescription for amputating external charged states. This issue is not always addressed explicitly in the literature~\cite{EffectiveMass_skalozub, SpectralProperties_ghosh, ChargedAndNeutral_liu}. For this reason, we derive the LSZ reduction formula for massive charged vector particles in the presence of a magnetic field. To the best of our knowledge, this constitutes a new result and further illustrates the usefulness of the Ritus eigenfunction representation in magnetic field background calculations.

This paper is organized as follows. In Sec.~\ref{sec.WMF}, the propagator of a massive charged vector boson in a homogeneous magnetic field is derived using the Ritus eigenfunction method after diagonalizing the equations of motion. In Sec.~\ref{sec.RitusToSchwinger}, the propagator is rewritten in the Schwinger proper-time representation by summing over Landau levels and performing the relevant integrations, where a discrepancy with previous results~\cite{ElectroweakProcesses_kuznetsov} is explicitly identified. In Sec.~\ref{sec.LSZMag}, the LSZ reduction formula for charged vector bosons in a magnetic background is obtained. Section~\ref{sec.concl} contains the conclusions.

\section{W Boson Propagator in an External Magnetic Field (Ritus Eigenfunction Method)} \label{sec.WMF}

\subsection{Model: Reduced electroweak Lagrangian}

For the sake of definiteness\footnote{As discussed in Ref.~\cite{ChargedMassive_iablokov}, several conventions and subtleties arise when formulating the electroweak Lagrangian after spontaneous symmetry breaking (SSB). These choices, together with issues related to gauge transformations and residual symmetries, are reviewed in Appendix~\ref{sec.Symmetry}.}, let us start by introducing the electroweak Lagrangian after SSB, restricting ourselves to the sector describing the interaction between the massive charged vector boson and the electromagnetic field~\cite{QuantumFieldTheory_schwartz}. 
The corresponding Lagrangian density, reads
\begin{equation} \label{eq:Lagrangian}
\begin{split}
\mathcal{L}_{W} & = - \frac{1}{4} F^{\mu \nu} F_{\mu \nu} - \frac{1}{2} W^{(+) \mu \nu} W_{\mu \nu}^{(-)} + m_{W}^{2} W^{(+)}_{\mu} W^{(-) \mu} \\
& + ie \left[ F^{\mu \nu} W_{\mu}^{(+)} W_{\nu}^{(-)} - W^{(+) \mu \nu} A_{\mu} W_{\nu}^{(-)} + W^{(-) \mu \nu} A_{\mu} W_{\nu}^{(+)} \right] + e^{2} \left[ A^{\mu} W_{\mu}^{(+)} A^{\nu} W_{\nu}^{(-)} - A^{\mu} A_{\mu} W^{(+)}_{\nu} W^{(-) \nu} \right],
\end{split}
\end{equation}
where $A^{\mu}$ denotes the photon field, $W^{(\pm)\mu}$ the charged massive vector boson fields,
\begin{equation}
F^{\mu \nu} \equiv \partial^{\mu} A^{\nu} - \partial^{\nu} A^{\mu},
\qquad
W^{(\pm)\mu \nu} \equiv \partial^{\mu} W^{(\pm)\nu} - \partial^{\nu} W^{(\pm)\mu},
\end{equation}
and $e$ is the electric charge. Note that Eq.~\ref{eq:Lagrangian} has a remaining global $U(1)_{em}$ symmetry.

By introducing the covariant derivative, 
\begin{equation}
    D^{\mu} \equiv \partial^{\mu} + ie A^{\mu},
\end{equation}
we  rewrite Eq.~\eqref{eq:Lagrangian} in a compact form, yielding
\begin{equation} \label{eq:Lagrangian2}
\begin{split}
\mathcal{L}_{W} & = - \frac{1}{4} F^{\mu \nu} F_{\mu \nu} - \frac{1}{2} \left( D^{\mu *} W^{(+) \nu} - D^{\nu *} W^{(+) \mu} \right) \left( D_{\mu} W^{(-)}_{\nu} - D_{\nu} W^{(-)}_{\mu} \right) + m_{W}^{2} W^{(+)}_{\mu} W^{(-) \mu} + ie F^{\mu \nu} W^{(+)}_{\mu} W_{\nu}^{(-)}.
\end{split}
\end{equation}
The equation of motion (EOM) for the $W^{-}$ field, obtained from Eq.~\eqref{eq:Lagrangian2}, reads 
\begin{equation} \label{eq:EOM0}
\begin{split}
& \Big[ \left( D^{2} + m_{W}^{2} \right) g^{\alpha \beta} - D^{\alpha} D^{\beta} + 2 i e F^{\alpha \beta} \Big] W_{\beta}^{(-)}(x) = 0,
\end{split}
\end{equation}
where $D^{2} \equiv D^{\mu} D_{\mu}$, the metric tensor
  $g^{\alpha \beta} = \mathrm{diag}(1,-1,-1,-1)$ , and the commutator
  \begin{equation}
  [ D_{\alpha}, D_{\beta} ] = i e F_{\alpha \beta}
  \end{equation}
has been used.

Note that in the case of a constant external electromagnetic field, there is an additional constrain, given by
\begin{equation}\label{eq:GaugeCondition}
D^{\mu} W_{\mu}^{(-)}(x) = 0,
\end{equation}
which is obtained  by applying a covariant derivative on Eq.~\eqref{eq:EOM0}. Thus, for a constant electromagnetic field, once we substitute Eq.~\eqref{eq:GaugeCondition} into Eq.~\eqref{eq:EOM0}, the EOM reduces to\footnote{The EOM for $W^{+}$ follows by charge conjugation.}
\begin{equation} \label{eq:EOM1}
\begin{split}
& \Big[\left( D^{2} + m_{W}^{2} \right) g^{\alpha \beta} + 2 i e F^{\alpha \beta} \Big] W_{\beta}^{(-)}(x) = 0,
\end{split}
\end{equation}
which is a tensor that has a non-diagonal matrix representation in Lorentz indices, making the analysis and physical interpretation a little bit involved, as noted previously in Ref.~\cite{ChargedMeson_scoccola}. To diagonalize the above equation it is convenient to perform a rotation in the Lorentz space using the transformation matrix ~\cite{NeutrinoSelfEnergy_elizalde} 
\begin{equation} \label{eq:RotMat}
N_{\alpha}{}^{\mu} \equiv \begin{pmatrix}
1 & 0 & 0 & 0 \\
0 & \frac{1}{\sqrt{2}} & \frac{-i}{\sqrt{2}} & 0 \\
0 & \frac{1}{\sqrt{2}} & \frac{i}{\sqrt{2}} & 0 \\
0 & 0 & 0 & 1
\end{pmatrix},
\end{equation}
satisfying
\begin{equation}
\left( N^{-1} \right)^{\mu}{}_{\alpha} = \left( N_{\alpha}{}^{\mu} \right)^{\dagger} \mbox{ and } N_{\alpha}{}^{\mu} \left( N^{-1} \right)_{\mu}{}^{\beta} = g_{\alpha}{}^{\beta},
\end{equation}
an approach that has been successfully employed in related contexts~\cite{MotionOfCharged_tsai}. Under this transformation, the electromagnetic field-strength tensor becomes diagonal,
\begin{equation}
\begin{split}
\tilde{F}_{\mu}{}^{\nu} \equiv N_{\mu}{}^{\alpha} \frac{i F_{\alpha}{}^{\beta}}{B} \left( N^{-1} \right)_{\beta}{}^{\nu} = \begin{pmatrix}
    0 & 0 & 0 & 0 \\
    0 & -1 & 0 & 0 \\
    0 & 0 & 1 & 0 \\
    0 & 0 & 0 & 0
\end{pmatrix}.
\end{split}
\end{equation}

The diagonalized EOM then reads\footnote{In Ref. \cite{MotionOfCharged_tsai}, a similar equation was solved as an eigenvalue problem.}
\begin{equation} \label{eq:EOM2}
\begin{split}
\Big[ \left( \hat{\tilde{\Pi}}^{2} - m_{W}^{2} \right) g_{\mu}{}^{\alpha} - 2 eB \tilde{F}_{\mu}{}^{\alpha} \Big] \tilde{W}_{\alpha}^{(-)}(x) = 0,
\end{split}
\end{equation}
where $\tilde{W}^{(-) \mu}(x) \equiv N^{\mu}{}_{\alpha}W^{(-) \alpha}(x)$ and $\hat{\tilde{\Pi}}^2={\hat{\tilde{\Pi}}_+}^{\! \! \mu} \, {\hat{\tilde{\Pi}}_-}_\mu$ are the transformed field and canonical momentum, respectively. Here, we used the notation  $\hat{\tilde{\Pi}}_{\pm}^\mu \equiv  i\tilde{D}^{\mu}_{\pm}$ with
\begin{equation}\label{eq:DerCov+-}
\begin{split}
\tilde{D}^{\mu}_{-} \equiv D^{\alpha} \left( N^{-1} \right)_{\alpha}{}^{\mu} \mbox{\ and \ } \tilde{D}^{\mu}_{+} \equiv N^{\mu}{}_{\alpha} D^{\alpha}.
\end{split}
\end{equation}
From now on, we denote all ``rotated'' quantities with $tilde$.

Introducing the spin-projection operator 
\begin{equation} \label{eq:MatProy}
\tilde{\Delta}^{\mu}{}_{\nu}(s_{3}) \equiv \begin{pmatrix}
    1 - s_{3}^{2} & 0 & 0 & 0 \\
    0 & s_{3} \frac{1 + s_{3}}{2} & 0 & 0 \\
    0 & 0 & - s_{3} \frac{1 - s_{3}}{2} & 0 \\
    0 & 0 & 0 & 1 - s_{3}^{2}
\end{pmatrix},
\end{equation}
with $s_{3}$ denoting the spin projection along the magnetic-field direction, and using
\begin{equation}
\sum_{s_{3} = \pm 1,0} \tilde{\Delta}^{\mu}{}_{\nu}(s_{3}) = g^{\mu}{}_{\nu}.
\end{equation}
we rewrite the EOM  in Eq.~\eqref{eq:EOM2} as
\begin{equation} \label{eq:EOM3}
\sum_{s_{3} = \pm 1, 0} \left( \hat{\tilde{\Pi}}^{2} - m_{W}^{2} + 2 e B s_{3} \right) \tilde{W}_{\mu}^{(-) s_{3}} (x) = 0,
\end{equation}
where $\tilde{W}^{(-) s_{3}}_{\mu} (x) \equiv \tilde{\Delta}_{\mu}{}^{\alpha} (s_{3}) \tilde{W}^{(-)}_{\alpha} (x)$ denotes the $s_3$ spin-projection of $\tilde{W}^{(-)}_{\alpha} (x)$. Eq.~\eqref{eq:EOM3} constitutes the starting point for the propagator construction presented in the following subsection.

\subsection{$W$ Boson Field Solution in a Magnetic Field}

To solve the equation of motion in Eq.~\eqref{eq:EOM3}, we must specify the form of the gauge field that gives rise to the constant external magnetic field. To describe a homogeneous magnetic field directed along the $z$ axis we adopt the antisymmetric (Landau) gauge $A^{\mu}(x) = \left( 0, 0, Bx^{1},0 \right)$ even so several gauge choices are possible. In this gauge, Eq.~\eqref{eq:EOM3} becomes
\begin{equation}\label{eq:EOM3lan}
\sum_{s_{3} = \pm1,0} \left[ i \partial^{0} i \partial_{0} + i \partial^{1} i \partial_{1} + i \left( \partial^{2} + i eB x^{1} \right) i \left( \partial_{2} + ieB x_{1} \right) + i \partial^{3} i \partial_{3} - m_{W}^{2} + 2eB s_{3} \right] \tilde{W}^{(-) s_{3}}_{\mu}(x) = 0.
\end{equation}

Since Eq.~\eqref{eq:EOM3lan} shows translational invariance in the $x^{0}$, $x^{2}$, and $x^{3}$ directions, we adopt the ansatz
\begin{equation} \label{eq:ansatz}
\begin{split}
\tilde{W}^{(-) s_{3}}_{\mu}(x) = e^{-i r \left( E x_{0} + p^{2} x_{2} + p^{3}x_{3} \right)} X^{r}(x^{1}) \tilde{\bar{\varepsilon}}^{(-)s_{3}}_{\mu}
\end{split}
\end{equation}
where $r=\pm1$ distinguishes between particle and antiparticle solutions, $X^{r}(x^{1})$ encodes the dependence on the transverse coordinate, and $\tilde{\bar{\varepsilon}}_{\mu}^{(-)s_{3}}$ are the projected polarization vectors, given by
\begin{equation}\label{polaproj}
    \tilde{\bar{\varepsilon}}_{\mu}^{(-)s_{3}}\equiv 
    {\tilde{\Delta}_\mu}{}^{\nu}(s_3)\tilde{\bar{\varepsilon}}^{(-)}_{\nu}.
\end{equation}

\noindent At this point, in Eq.~\eqref{polaproj}, the explicit form and dependence of the polarization vectors is unknown, however, this will be solved and shown later on the present work.

Once we substitute the ansatz from Eq.~\eqref{eq:ansatz} into Eq. (\ref{eq:EOM3}), we get
\begin{equation} \label{eq:Weber}
\Bigg\{ \frac{\partial^{2}}{\partial \rho_{r}^{2}} + \Bigg[ - \frac{1}{4} \rho^{2}_{r} + \frac{1}{2} + n \Bigg] \Bigg\} X^{r}(\rho_{r}) = 0
\end{equation}
which is Weber’s differential equation~\cite{ACourseOfModernAnalysis_whittaker}. To obtain the above equation, we made the change of variable 
\begin{equation}\label{eq:CambioVariable}
\rho_{r} \equiv \sqrt{2|eB|} \left( p^{2}/eB - r x^{1} \right)
\end{equation} and defined the integer $n \equiv \ell-1 + \hat{s}_{3}$ to label the excitation energy level with $n=0, 1, 2, \ldots $~\cite{NeutrinoSelfEnergy_elizalde, StrongField_hattori}. Here,
\begin{equation}
\ell \equiv \frac{E^{2} - \left( p^{3} \right)^{2} - m_{W}^{2}}{2|eB|} + \frac{1}{2},
\end{equation}
is the Landau-level index, and $\hat{s}_{3} \equiv s_{eB} s_{3}$ with $s_{eB} \equiv \text{sign} (eB)$ the sign function.
  
The solutions of Eq. (\ref{eq:Weber}) are the parabolic cylinder functions \cite{ACourseOfModernAnalysis_whittaker}, 
\begin{equation}
    X^{r}_{n}(\rho_{r}) = N_{n} D_{n}(\rho_{r}),
\end{equation}
with normalization constant $N_{n} = \left( \sqrt{4 \pi |eB|}/(n!) \right)^{\frac{1}{2}}$. With the above solutions, we have fully determined the space-time dependence of the wave functions; however, the explicit form of the polarization vectors remains to be solved. To figure out the form of these vectors, we use Eq.~(\ref{eq:GaugeCondition}) and rewrite it as follows
\begin{equation} \label{eq:GaugeCond2}
\begin{split}
& \hat{\tilde{\Pi}}^{\mu}_{-}(x) \left( \tilde{\mathbb{E}}^r_{\ell, 1} \right)_{\! \!  \mu}^{\, \nu} \!  (p, x) \tilde{\bar{\varepsilon}}^{(-)}_{\nu} = 0, 
\end{split}
\end{equation}
where Eqs.~\eqref{eq:ansatz} and~\eqref{polaproj} were used and
\begin{equation}
     \left( \tilde{\mathbb{E}}^r_{\ell, c} \right)_{\! \! \mu}^{\, \nu} \! (p, x) \equiv \sum_{s_{3} \equiv \pm 1, 0} e^{-i r \left( E x_{0} + p^{2} x_{2} + p^{3}x_{3} \right)}  X^{r}_{\ell - 1 + c \hat{s}_{3}}(\rho_{r}) {\tilde{\Delta}_\mu}{}^{\nu}(s_3)
\end{equation}
was introduced with $c \equiv \{0, 1\}$.

Now, using the explicit form of $\hat{\tilde{\Pi}}^{\mu}_{\pm}(x)$ and taking into account the identities~\cite{ACourseOfModernAnalysis_whittaker}
\begin{equation} \label{eq:ParaCilDeriv}
\begin{split}
& \frac{\partial}{\partial z} D_{n}(z) = n D_{n-1}(z) - \frac{z}{2} D_{n}(z), \\
& n D_{n-1}(z) = z D_{n}(z) - D_{n+1}(z),
\end{split}
\end{equation}
then, the action of the differential operators on the parabolic cylinder functions in Eq.~\eqref{eq:GaugeCond2} can be easily evaluated, resulting in 
\begin{equation} \label{eq:transv}
\left( \tilde{\mathbb{E}}^r_{\ell, 0} \right)_{\! \! \mu}^{\, \nu} \! (p, x) \, \tilde{\bar{p}}^{*\mu} \tilde{\bar{\varepsilon}}^{(-)}_\nu = 0,
\end{equation}
with
\begin{equation} \label{eq:pmu}
\begin{split}
\tilde{\bar{p}}^{\mu *} & \equiv \left(E, is_{eB} \sqrt{|eB|} \sqrt{\ell_{+}}, -is_{eB} \sqrt{|eB|} \sqrt{\ell_{-}}, p^{3} \right)
\end{split}
\end{equation}
plays the role of an effective four-momentum in the presence of the magnetic field, valid for $\ell \geq 1$. In obtaining this result, we have defined $\sqrt{\ell_{\pm}} \equiv \sqrt{\ell - 1 + s_{\pm}}$, with $s_{\pm} \equiv \left( 1 \pm s_{eB} \right) / 2$ and used the relation between $n$ and $\ell$ below Eq.~\eqref{eq:CambioVariable}. The effect of the differential operators over the Ritus eigenfunctions

\begin{equation}
\begin{split}
& \hat{\tilde{\Pi}}^{\mu}_{-}(x) \left( \tilde{\mathbb{E}}^r_{\ell, 1} \right)_{\! \! \mu}^{\, \nu} \! (p, x) =  \tilde{\bar{p}}^{*\mu} \left( \tilde{\mathbb{E}}^r_{\ell, 0} \right)_{\! \! \mu}^{\, \nu} \! (p, x) \\
& \hat{\tilde{\Pi}}^{\mu}_{+}(x) \left( \tilde{\mathbb{E}}^r_{\ell, 0} \right)_{\! \! \mu}^{\, \nu} \! (p, x) =  \tilde{\bar{p}}^{\mu} \left( \tilde{\mathbb{E}}^r_{\ell, 1} \right)_{\! \! \mu}^{\, \nu} \! (p, x)
\end{split}
\end{equation}

will be of relevance throughout the present work. It is worth emphasizing that the solutions to the EOM obtained here exhibit intrinsically quantum features, as reflected in the discrete energy spectrum arising from their dependence on the Landau level.

\subsection{Calculation of the Polarization Vectors in a Constant Magnetic Field}

Since the knowledge of the magnetic four-momentum in Eq.~\eqref{eq:pmu} is not sufficient to fully determine the polarization vectors $\tilde{\bar{\varepsilon}}^{(-)}_{\nu}$ appearing in Eq.~\eqref{eq:transv}, we need additional constraints. To this end, we use the conserved current associated with the Lagrangian in Eq.~\eqref{eq:Lagrangian},
\begin{equation} \label{eq:CorrienteConservadaGen}
J_{\mu} \left( W^{(-)}_{\mathbf{\lambda}}(x), W^{(+)}_{\mathbf{\lambda^{\prime}}}(x) \right) = \frac{\partial \mathcal{L}}{\partial \left( \partial^{\mu} W_{\nu, \mathbf{\lambda^{\prime}}}^{(+)} \right)} \left( -i W_{\nu, \mathbf{\lambda^{\prime}}}^{(+)} \right) + \frac{\partial \mathcal{L}}{\partial \left( \partial^{\mu} W^{(-)}_{\nu, \mathbf{\lambda}} \right)} \left( i W^{(-)}_{\nu, \mathbf{\lambda}} \right),
\end{equation}
satisfying
\begin{equation} \label{def:ortonor}
\int dx^{1} J_{0} \left( W^{(-)}_{\mathbf{\lambda}}(x), W^{(+)}_{\mathbf{\lambda^{\prime}}}(x) \right) = 4 \pi E \delta_{\mathbf{\lambda, \lambda^{\prime}}},
\end{equation}
which provides us with an independent condition on the polarization structure\footnote{This approach was previously employed in Ref.~\cite{VectorBoson_nikishov}.}. The bold indices $\lambda, \lambda^{\prime}$ label the polarization states, and no summation over these repeated indices is implied. Together with the transversality condition in Eq.~\eqref{eq:transv}, this relation allows for a complete determination of the polarization vectors.

As shown before, the Landau-level index $\ell$ is directly related to the energy spectrum of the system. This observation is particularly important because the number of admissible spin projection configurations depends on the Landau level, as discussed in Ref.~\cite{StrongField_hattori}. In the presence of a magnetic field, the LLL admits a single spin projection for a spin-1 particle, while at higher Landau levels one or two additional spin projections become allowed, increasing the number of independent polarization states.

Using Eqs.~\eqref{eq:transv} and~\eqref{def:ortonor}, the polarization vectors can be explicitly constructed for each value of $\ell$, for a given polarization $\mathbf{\lambda}$ and a magnetic four-momentum $\tilde{\bar{p}}$. In what follows, we show the polarization vectors we obtained; for details see Appendix~\ref{sec.PolVec}.  

\subsubsection{Lowest Landau level (LLL)}
For the LLL, $\ell = 0$, only one polarization vector exists,
\begin{equation} \label{eq:eBGen0}
\tilde{\bar{\varepsilon}}^{(-) \mu}_{\mathbf{1}}(\tilde{\bar{p}}, \ell = 0) = \left(0, s_{+}, s_{-}, 0 \right),
\end{equation}
where $s_{\pm} \equiv (1 \pm s_{eB})/2$ and the polarization subindex $\lambda = \mathbf{1}$ emphasizes that this is the only polarization vector possible in this energy state.

\subsubsection{First excited Landau level}
For the excited Landau level with $\ell = 1$, two independent polarization vectors are available,
\begin{equation}
\begin{split}
\tilde{\bar{\varepsilon}}^{(-) \mu}_{\mathbf{1}}(\tilde{\bar{p}}, \ell = 1) & = \mathcal{N}_{1,1} \left( p^{3}, 0, 0, E \right), \\
\tilde{\bar{\varepsilon}}^{(-) \mu}_{\mathbf{2}} (\tilde{\bar{p}}, \ell = 1) & =  \mathcal{N}_{2,1} \left(E, \frac{-i m_{\bot}^{2}}{\sqrt{|eB|}} s_{+}, \frac{-i m_{\bot}^{2}}{\sqrt{|eB|}} s_{-}, p^{3} \right),
\end{split}
\end{equation}
with
\begin{equation}
    \mathcal{N}_{1,\ell}=\sqrt{\frac{1}{m_{\bot}^{2}}} \mbox{\ and \ }
    \mathcal{N}_{2,\ell}= \sqrt{\frac{1}{m_{\bot}^{2}}} \sqrt{\frac{|eB|\ell}{m_{\bot}^{2} - |eB|\ell}},
\end{equation}
the normalization factors, and the notation $m_{\bot}^{2} \equiv E^{2} - \left( p^{3} \right)^{2}$ is used~\cite{ChargedMeson_scoccola}.

\subsubsection{Higher excited Landau levels}
For higher Landau levels, $\ell > 1$, three linearly independent polarization vectors exist,
\begin{equation} \label{eq:VecsPol}
\begin{split}
\tilde{\bar{\varepsilon}}^{(-) \mu}_{\mathbf{1}}(\tilde{\bar{p}}, \ell > 1) & = \mathcal{N}_{1,\ell} \left( p^{3}, 0, 0, E \right), \\
\tilde{\bar{\varepsilon}}^{(-) \mu}_{\mathbf{2}}(\tilde{\bar{p}}, \ell > 1) & = \mathcal{N}_{2,\ell} \left(E, \frac{-i m_{\bot}^{2}}{\sqrt{|eB|\ell}} s_{+}, \frac{-i m_{\bot}^{2}}{\sqrt{|eB|\ell}} s_{-}, p^{3} \right), \\
\tilde{\bar{\varepsilon}}^{(-) \mu}_{\mathbf{3}}(\tilde{\bar{p}}, \ell > 1) & = \mathcal{N}_{3,\ell} 
\Big(E, -i \sqrt{|eB|\ell} s_{+} + i \frac{m_{\bot}^{2} - |eB|\ell}{\sqrt{|eB|(\ell-1)}} s_{-}, -i \sqrt{|eB|\ell} s_{-} + i \frac{m_{\bot}^{2} - |eB|\ell}{\sqrt{|eB|(\ell-1)}} s_{+}, p^{3} \Big).
\end{split}
\end{equation}
with 
\begin{equation}
   \mathcal{N}_{3,\ell} = \left\{|eB| \ell - m_{\bot}^{2} + \left[ \frac{m_{\bot}^{2} - |eB|\ell}{\sqrt{|eB|(\ell-1)}} \right]^{2} \right\}^{- \frac{1}{2}}.
\end{equation}

In this case, the sum over physical polarizations satisfies
\begin{equation} \label{eq:SumPolMag}
\begin{split}
& \sum_{\lambda = 1}^{3} \tilde{\bar{\varepsilon}}^{(-) \mu}_{\mathbf{\lambda}} (\tilde{\bar{p}}, \ell > 1) \tilde{\bar{\varepsilon}}^{(-) \nu *}_{\mathbf{\lambda}} (\tilde{\bar{p}}, \ell > 1) = - g^{\mu \nu} + \frac{\tilde{\bar{p}}^{\mu} \tilde{\bar{p}}^{\nu *}}{m_{W}^{2}}, \\
& \sum_{\lambda = 1}^{3} \tilde{\bar{\varepsilon}}^{(+) \mu}_{\mathbf{\lambda}} (\tilde{\bar{p}}, \ell > 1) \tilde{\bar{\varepsilon}}^{(+) \nu *}_{\mathbf{\lambda}}(\tilde{\bar{p}}, \ell > 1) = - g^{\mu \nu} + \frac{\tilde{\bar{p}}^{\mu *} \tilde{\bar{p}}^{\nu}}{m_{W}^{2}},
\end{split}
\end{equation}
where $\mathbf{\lambda}$ labels the polarization states, and $\tilde{\bar{\varepsilon}}^{(+)\nu}_{\mathbf{\lambda}}$ denotes the polarization vectors of the $W^{+}$ boson.

At this point, all ingredients entering the ansatz in Eq.~\eqref{eq:ansatz} have been fully specified. The resulting solutions describe charged massive vector bosons in a constant magnetic field. In the next step, these fields will be promoted to quantum operators via canonical quantization, allowing for the construction of the corresponding propagator.

\subsection{Calculation of the Massive Charged Vector Boson Propagator in a Magnetic Field}

Following the standard canonical quantization procedure~\cite{QuantumElectrodynamicsStrong_greiner}, we promote the solutions obtained above to quantum field operators. In the Ritus basis, the quantized fields are expanded as
\begin{equation} \label{eq:CuantizacionMag}
\begin{split}
\hat{\tilde{W}}^{(-) \mu}(x) & = \SumInt \frac{d^{3}p}{(2 \pi)^{3}} \frac{1}{\sqrt{2E_{\tilde{\bar{p}}}}} \sum_{\lambda=1}^{3} \left( \hat{a}_{\tilde{\bar{p}}, \lambda} \left( \tilde{\mathbb{E}}^{+}_{\ell, 1} \right)^{\! \mu}_{\, \nu} \! (p, x) \,  \tilde{\bar{\varepsilon}}^{(-) \nu}_{\mathbf{\lambda}}(\tilde{\bar{p}}, \ell) + \hat{b}_{\bar{p}, \lambda}^{\dagger} \left( \tilde{\mathbb{E}}^{-}_{\ell, 1} \right)^{\! \mu}_{\, \nu} \! (p, x) \,  \tilde{\bar{\varepsilon}}_{\mathbf{\lambda}}^{(+) \nu *}(\tilde{\bar{p}}, \ell)  \right), \\
\hat{\tilde{W}}^{(+) \mu}(x) & = \SumInt \frac{d^{3}p}{(2 \pi)^{3}} \frac{1}{\sqrt{2E_{\tilde{\bar{p}}}}} \sum_{\lambda = 1}^{3} \left( \hat{a}^{\dagger}_{\tilde{\bar{p}}, \lambda} \left(\tilde{\mathbb{E}}^{+ *}_{\ell , 1} \right)^{\! \mu}_{\, \nu} \! (p, x) \,  \tilde{\bar{\varepsilon}}_{\mathbf{\lambda}}^{(-) \nu *}(\tilde{\bar{p}}, \ell) + \hat{b}_{\tilde{\bar{p}}, \lambda} \left(\tilde{\mathbb{E}}^{- *}_{\ell , 1} \right)^{\! \mu}_{\, \nu} \! (p, x) \, \tilde{\bar{\varepsilon}}_{\mathbf{\lambda}}^{(+) \nu}(\tilde{\bar{p}}, \ell)  \right),
\end{split}
\end{equation}
with the notation
\begin{equation}
\SumInt \frac{d^{3}p}{(2 \pi)^{3}} \equiv \frac{1}{2 \pi} \sum_{\ell = 0}^{\infty}  \int_{-\infty}^{\infty} \frac{dp^{2}}{2 \pi} \int_{-\infty}^{\infty} \frac{dp^{3}}{2 \pi},
\end{equation}
\noindent where the sum over all Landau levels and corresponding available polarization vectors is taken into account.

The creation and annihilation operators are imposed to satisfy,
\begin{equation} \label{eq:ConmutacionMag}
    \left[ \hat{a}_{\tilde{\bar{p}}, \lambda}, \hat{a}^{\dagger}_{\tilde{\bar{p}}^{\prime}, \lambda^{\prime}} \right] = \left( 2 \pi \right)^{3} \delta_{\ell, \ell'} \delta_{\lambda, \lambda'} \delta \left( p_{2} - p'_{2} \right) \delta \left( p_{3} - p'_{3} \right) = \left[ \hat{b}_{\tilde{\bar{p}}, \lambda}, \hat{b}^{\dagger}_{\tilde{\bar{p}}^{\prime}, \lambda^{\prime}} \right],
\end{equation}
with all other commutators vanishing.

To calculate the $W$-boson Feynman propagator, we use the standard definition\cite{QuantumFieldTheory_schwartz}
\begin{equation} \label{eq:ProdTempOrd}
\begin{split}
\tilde{D}^{\mu \nu}_{B}(x,y) & \equiv \bra{0} \mathcal{T} \left( \hat{\tilde{W}}^{(-) \mu}(x) \hat{\tilde{W}}^{(+) \nu}(y) \right) \ket{0} \\
& = \Big\langle \hat{\tilde{W}}^{(-) \mu}(x) \hat{\tilde{W}}^{(+) \nu}(y) \Big\rangle \theta \left( x^{0} - y^{0} \right) + \Big\langle \hat{\tilde{W}}^{(-) \nu}(y) \hat{\tilde{W}}^{(+) \mu}(x) \Big\rangle \theta\left( y^{0} - x^{0} \right),
\end{split}
\end{equation}
where $\mathcal{T}$ denotes the time-ordered product. Using the mode expansion~\eqref{eq:CuantizacionMag} and the polarization sum in Eq.~\eqref{eq:SumPolMag}, the first term in Eq.~\eqref{eq:ProdTempOrd}, becomes
\begin{equation} \label{eq:FeynProp1st}
\begin{split}
\langle \hat{\tilde{W}}^{(-) \mu}(x)  \hat{\tilde{W}}^{(+) \nu}(y) \rangle  
 & = \SumInt \frac{d^{3}p}{(2 \pi)^{3}} \frac{1}{2E_{\tilde{\bar{p}}}} \left( \tilde{\mathbb{E}}^{+}_{\ell, 1} \right)^{\! \mu}_{\, \nu} \! (p, x) \,  \left( -g^{\alpha \beta} + \left( 1 - \delta_{\ell, 0} \right) \frac{\tilde{\bar{p}}^{\alpha} \tilde{\bar{p}}^{\beta *}}{m_{W}^{2}} \right) \left( \tilde{\mathbb{E}}^{+ *}_{\ell, 1} \right)_{\! \! \beta }^{\, \nu} \!(p, y).
\end{split}
\end{equation}
which, by recalling the action of the operator $\hat{\tilde{\Pi}}^{\alpha}_{+}(x) \left( \tilde{\mathbb{E}}^{+}_{\ell , 0} \right)_{\! \! \alpha}^{\, \mu} \! (p, x) = \tilde{\bar{p}}^{\alpha} \left( \tilde{\mathbb{E}}^{+}_{\ell, 1} \right)_{\! \alpha}^{\, \mu} \! (p, x)$, 
allows us to rewrite Eq.~\eqref{eq:FeynProp1st} as
\begin{equation} \label{eq:PropRitus1}
\begin{split}
&\langle \hat{\tilde{W}}^{(-) \mu}(x)  \hat{\tilde{W}}^{(+) \nu}(y) \rangle \\
& = \SumInt \frac{d^{3}p}{(2 \pi)^{3}} \frac{1}{2E_{\tilde{\bar{p}}}} \Bigg[ \left( - g^{\alpha \beta} \right) \left( \tilde{\mathbb{E}}^{+}_{\ell , 1} \right)^{\! \mu}_{\, \alpha} \! (p, x) \, \left( \tilde{\mathbb{E}}^{+ *}_{\ell, 1} \right)_{\! \! \beta}^{\, \nu} \! (p, y) + \left( 1 - \delta_{\ell, 0} \right) \left( \frac{\hat{\tilde{\Pi}}^{\alpha}_{+}(x) \hat{\tilde{\Pi}}^{\beta *}_{+}(y)}{m_{W}^{2}} \right)  \left( \tilde{\mathbb{E}}^{+}_{\ell , 0} \right)^{\! \mu}_{\, \alpha} \! (p, x) \, \left( \tilde{\mathbb{E}}^{+ *}_{ \ell, 0} \right)^{\, \nu}_{\! \! \beta} \! (p, y) \Bigg],
\end{split}
\end{equation}
where the operator action is understood to act on the corresponding Ritus eigenfunctions.

A similar treatment for the second term in Eq.~(\ref{eq:ProdTempOrd}), yields
\begin{equation} \label{eq:PropRitus2}
\begin{split}
&\Big\langle \hat{\tilde{W}}^{(+) \nu}(y) \hat{\tilde{W}}^{(-) \mu}(x) \Big\rangle \\
& = \SumInt \frac{d^{3}p}{(2 \pi)^{3}} \frac{1}{2 E_{\tilde{\bar{p}}}} 
\Bigg[ \left( - g^{\alpha \beta} \right) \left( \tilde{\mathbb{E}}^{-}_{\ell , 1} \right)^{\! \mu}_{\, \alpha} \! (p, x) \,  \left( \tilde{\mathbb{E}}^{- *}_{\ell , 1} \right)_{\! \! \beta}^{\, \nu} \! (p, y) + \left( 1 - \delta_{\ell, 0} \right) \left( \frac{\hat{\tilde{\Pi}}^{\alpha}_{+}(x) \hat{\tilde{\Pi}}^{\beta *}_{+}(y)}{m_{W}^{2}} \right)  \left( \tilde{\mathbb{E}}^{-}_{ \ell, 0} \right)^{\! \mu}_{\, \alpha} \! (p, x) \, \left( \tilde{\mathbb{E}}^{- *}_{\ell , 0} \right)_{\! \! \beta}^{\, \nu} \! (p, y) \Bigg] \Bigg|_{\substack{p^{2} \rightarrow -p^{2} \\ p^{3} \rightarrow -p^{3}}},
\end{split}
\end{equation}
where an additional change of variables, $p^{2} \rightarrow - p^{2}$ and $p^{3} \rightarrow - p^{3}$, is indicated in order to combine Eq.~\eqref{eq:PropRitus2} with Eq.~\eqref{eq:PropRitus1}. Thus, once we substitute Eqs.~\eqref{eq:PropRitus1} and~\eqref{eq:PropRitus2} into Eq.~\eqref{eq:ProdTempOrd}, and use the integral representation for the Heaviside theta function~\cite{Gradshteyn:1943cpj}
\begin{equation}
\theta \left( x_{0} - y_{0} \right) = \lim_{\varepsilon \rightarrow 0} \frac{1}{2 \pi i} \int^{\infty}_{-\infty} \frac{e^{i \left( x_{0} - y_{0} \right) \omega}}{\omega -i \varepsilon} d \omega,
\end{equation}
we obtain 
\begin{equation} \label{eq:PropagadorMagnetico}
\begin{split}
\tilde{D}^{\mu \nu}_{B}(x,y) & = i \SumInt \frac{d^{4}p}{(2 \pi)^{4}} \frac{\left( \tilde{\mathbb{E}}^{+}_{\ell , 1} \right)^{\! \mu}_{\, \alpha} \! (p, x) \,  \left( -g^{\alpha \beta} + \left( 1 - \delta_{\ell, 0} \right) \frac{\tilde{\mathscr{P}}^{\alpha} \tilde{\mathscr{P}}^{\beta *}}{m_{W}^{2}} \right) \left( \tilde{\mathbb{E}}^{+ *}_{\ell , 1}  \right)_{\! \! \beta}^{\, \nu} \! (p, y)}{\tilde{\mathscr{P}}^{2} - m_{W}^{2} + i \varepsilon},
\end{split}
\end{equation}
the propagator for a charged vector boson in the Ritus eigenfunction representation, where 
\begin{equation}
\mathscr{\tilde{P}}^{\mu *} \equiv \left( p^{0}, is_{eB} \sqrt{|eB|} \sqrt{\ell_{+}}, -is_{eB} \sqrt{|eB|} \sqrt{\ell_{-}}, p^{3} \right)
\end{equation} with $\mathscr{\tilde{P}}^{2} \equiv \mathscr{\tilde{P}}^{\mu} \mathscr{\tilde{P}}^{*}_{\mu}$ and 
\begin{equation}
    \SumInt \frac{d^{4}p}{(2 \pi)^{4}}\equiv \frac{1}{2 \pi} \sum_{\ell = 0}^{\infty}  \int_{-\infty}^{\infty}\frac{dp^{0}}{2 \pi}\int_{-\infty}^{\infty} \frac{dp^{2}}{2 \pi} \int_{-\infty}^{\infty} \frac{dp^{3}}{2 \pi}.
\end{equation}

In Eq.~\eqref{eq:PropagadorMagnetico}, the $W$-boson propagator in a constant magnetic field is displayed in the Ritus representation and the unitary gauge. The analytic structure of this propagator closely parallels the vacuum case, differing only through the replacement of plane waves by Ritus eigenfunctions and the four-momenta by the momenta $\tilde{\mathscr{P}}^{\mu}$~\cite{QuantumFieldTheory_schwartz}. This formulation has been used in related contexts~\cite{NeutrinoSelfEnergy_elizalde} and will be particularly advantageous in Sec.~\ref{sec.LSZMag}, where the LSZ reduction procedure in a magnetic background is developed.

The Ritus representation for the propagator is conceptually transparent; however, it is not the most commonly employed method in practical calculations compared to the Schwinger proper-time parametrization. For this reason, in the next section, we present a novel methodology to translate the propagator from the Ritus basis to the Schwinger representation. This connection also allows us to explicitly exhibit a slight discrepancy between the expression obtained here and the result reported in Ref.~\cite{ElectroweakProcesses_kuznetsov}.

\section{From Ritus Eigenfunctions to Schwinger Parameters} \label{sec.RitusToSchwinger}

In order to make contact with the $W$-boson propagator in a constant magnetic field background in the Schwinger proper time representation~\cite{OnGaugeInvariance_schwinger}, let us start by rewriting Eq.~\eqref{eq:PropagadorMagnetico} as
\begin{equation}\label{eq:proptimewprop}
\begin{split}
\tilde{D}^{\mu \nu}_{B}(x,y) & = \int_{-\infty}^{\infty} \frac{dp^{0} \, dp^{2} \, dp^{3}}{(2 \pi)^{3}} \int_{0}^{\infty} \frac{ds}{2 \pi} \; \sum_{\ell = 0}^{\infty} e^{-i s \left[ m^{2}_{W} + (p^{3})^{2} - (p^{0})^{2} + 2  |eB| \left( \ell - 1 \right) \right] } \\
& \times \Bigg[ \left( -g^{\alpha \beta} \right) \sum_{s_{3}, s_{3}^{\prime}} X^{+}_{\ell - 1 + \hat{s}_{3}}(\rho_{+}) \tilde{\Delta}^{\mu}{}_{\alpha}(s_{3}) X^{+}_{\ell - 1 + \hat{s}_{3}^{\prime}}(\rho_{+}^{\prime}) \tilde{\Delta}^{\nu}{}_{\beta}(s_{3}^{\prime}) \\
& + \left( 1 - \delta_{\ell, 0} \right) \frac{\hat{\tilde{\Pi}}^{\alpha}_{+}(x) \hat{\tilde{\Pi}}^{\beta *}_{+}(y)}{m_{W}^{2}} \sum_{s_{3}, s_{3}^{\prime}} X^{+}_{\ell - 1}(\rho_{+}) \tilde{\Delta}^{\mu}{}_{\alpha}(s_{3}) X^{+}_{\ell - 1}(\rho_{+}^{\prime}) \tilde{\Delta}^{\nu}{}_{\beta}(s_{3}^{\prime}) \Bigg] e^{-i \left[ p^{0} \left( x_{0} - y_{0} \right) + p^{2} \left( x_{2} - y_{2} \right) + p^{3} \left( x_{3} - y_{3} \right) \right]}.
\end{split}
\end{equation}
where the identity
\begin{equation}\label{eq:Schwinger}
\frac{1}{x} = -i \int_{0}^{\infty} ds \hspace{0.1cm} e^{isx}, \hspace{1cm} \textrm{if} \hspace{0.2cm} \textrm{Im}(x) > 0,
\end{equation}
was used to turn the denominator into an exponential factor and the variable $\rho_{+}$ is defined in Eq.~\eqref{eq:CambioVariable} for $x$ and the prime, $\rho'_{+}$, is used for $y$.

The sum of Landau-levels can be easily performed by taking into account the integral representation of the parabolic cylinder function \cite{Bateman:100233}
\begin{equation}\label{parabintrepp}
\begin{split}
D_{n}(z) = \left( \frac{2}{\pi} \right)^{\frac{1}{2}} e^{\frac{z^{2}}{4}} \int^{\infty}_{0} dt \; e^{- \frac{t^{2}}{2}} t^{n} \cos{\left( zt - \frac{n \pi}{2} \right)},
\end{split}
\end{equation}
together with the identity~\cite{Gradshteyn:1943cpj} 
\begin{equation}\label{identsumparab}
\begin{split}
& \sum_{\ell = 0}^{\infty} e^{-is \left( 2 |eB| \ell \right)} \frac{(-i r t)^{\ell} (-i r^{\prime} t^{\prime})^{\ell}}{\ell !} = \exp \left\{ - e^{-2i |eB| s} r t r^{\prime} t^{\prime} \right\}. \\
\end{split}
\end{equation}
Thus, using Eqs.\eqref{parabintrepp} and~\eqref{identsumparab} in Eq.~\eqref{eq:proptimewprop}, we obtain
\begin{equation} \label{eq:PreSubsecs}
\begin{split}
\tilde{D}^{\mu \nu}_{B}(x,y) & = \int_{-\infty}^{\infty} \frac{dp^{0} \, dp^{2} \, dp^{3}}{(2 \pi)^{3}} \int_{0}^{\infty} \frac{ds}{2 \pi} \; e^{-i s \left[ m^{2}_{W} + (p^{3})^{2} - (p^{0})^{2} \right] } \sqrt{\frac{-2i \pi |eB|}{\sin{(2 |eB|s)}}} \\
& \times \Bigg[ \left( -g^{\alpha \beta} \right) \sum_{s_{3},s_{3}^{\prime}} \tilde{\Delta}^{\mu}{}_{\alpha}(s_{3}) \tilde{\Delta}^{\nu}{}_{\beta}(s_{3}^{\prime}) e^{2i eB s_{3} s} + \frac{\hat{\tilde{\Pi}}^{\alpha}_{+}(x) \hat{\tilde{\Pi}}^{\beta *}_{+}(y)}{m_{W}^{2}} \sum_{s_{3}, s_{3}^{\prime}} \tilde{\Delta}^{\mu}{}_{\alpha}(s_{3}) \tilde{\Delta}^{\nu}{}_{\beta}(s_{3}^{\prime}) \Bigg] \\
& \times \exp{\left\{i \frac{\left( \rho_{+}^{2} + \rho_{+}^{\prime 2} \right) \cos{\left( 2 |eB|s \right) - 2 \rho_{+} \rho_{+}^{\prime}}}{4 \sin{\left( 2 |eB|s \right)}} \right\}} e^{-i \left[ p^{0} \left( x_{0} - y_{0} \right) + p^{2} \left( x_{2} - y_{2} \right) + p^{3} \left( x_{3} - y_{3} \right) \right]}.
\end{split}
\end{equation}
Note that Eq.~\eqref{eq:PreSubsecs} makes it clear that the dependence on the spin degrees of freedom is entirely encoded in the tensor structure and in the action of the differential operators.

In the above equation, once the exponential argument within curly brackets is expanded, it is not hard to see that it has a Gaussian form in the momentum component $p^{2}$; thus,  once this integral is performed in Eq.~\ref{eq:PreSubsecs}, it reduces to
\begin{equation}\label{eq:PreSubsecscovv}
\begin{split}
\tilde{D}^{\mu \nu}_{B}(x,y) & = \int \frac{dp^{0}}{2 \pi} \int \frac{dp^{3}}{2 \pi} \int^{\infty}_{0} \frac{ds}{2 \pi} e^{-is \left[ -p_{\parallel}^2 + m_{W}^{2} \right]} \frac{-ieB}{2 \sin{(eB s)}} \\
& \times \Bigg[ \left( -g^{\alpha \beta} \right) \sum_{s_{3}} \tilde{\Delta}^{\mu}{}_{\alpha}(s_{3}) \tilde{\Delta}^{\nu}{}_{\beta}(s_{3}) e^{2i eB s_{3} s} + \frac{\hat{\tilde{\Pi}}^{\alpha}_{+}(x) \hat{\tilde{\Pi}}^{\beta *}_{+}(y)}{m_{W}^{2}} \sum_{s_{3}, s_{3}^{\prime}} \tilde{\Delta}^{\mu}{}_{\alpha}(s_{3}) \tilde{\Delta}^{\nu}{}_{\beta}(s_{3}^{\prime}) \Bigg] \\
& \times \Omega(x,y) e^{-i  p\cdot\left( x - y \right)_{\parallel}} e^{\frac{ieB}{4} \cot{(eBs)} \left[ \left( x - y \right)_{\mu} \varphi^{\mu \sigma} \varphi_{\sigma}{}^{\nu} \left( x - y \right)_{\nu} \right]},
\end{split}
\end{equation}
\noindent where
\begin{equation}
\Omega(x,y) = e^{i \frac{eB}{2} \left( x_{\mu} \varphi^{\mu \nu} y_{\nu} + \frac{A_{\mu} (x)}{B} \varphi^{\mu \nu} \varphi_{\nu}{}^{\alpha} x_{\alpha} - \frac{A_{\mu} (y)}{B} \varphi^{\mu \nu} \varphi_{\nu}{}^{\alpha} y_{\alpha} \right)},
\end{equation} 
is the well-known Schwinger phase that encodes the propagator gauge dependence~\cite{ChargedMeson_scoccola}. In Eq.~\eqref{eq:PreSubsecscovv}, we have introduced the covariant notation
\begin{equation}
\begin{split}
p^{0} \left( x_{0} - y_{0} \right) + p^{3} \left( x_{3} - y_{3} \right) &\equiv p\cdot \left( x - y \right)_{\parallel}, \\
 \left( x_{1} + y_{1} \right) \left( x_{2} - y_{2} \right) &= \left( x_{\mu} \varphi^{\mu \nu} y_{\nu} + \frac{A_{\mu} (x)}{B} \varphi^{\mu \nu} \varphi_{\nu}{}^{\alpha} x_{\alpha} - \frac{A_{\mu} (y)}{B} \varphi^{\mu \nu} \varphi_{\nu}{}^{\alpha} y_{\alpha} \right) 
, \\
\left( x_{1} - y_{1} \right)^{2} + \left( x_{2} - y_{2} \right)^{2}
&= \left( x - y \right)_{\mu} \varphi^{\mu \sigma} \varphi_{\sigma}{}^{\nu} \left( x - y \right)_{\nu},
\end{split}
\end{equation}
where $\varphi^{\mu \nu} \equiv F^{\mu \nu} / B $ is the normalized electromagnetic field-strength tensor, and $\tilde{\varphi}^{\mu \nu} \equiv \tilde{F}^{\mu \nu} / B $ denotes its dual. 

A straightforward calculation of the operators $\hat{\tilde{\Pi}}^{\alpha}_{+}(x) \hat{\tilde{\Pi}}^{\beta *}_{+}(y)$ acting on Schwinger's phase and the exponential factor allows us to rewrite Eq.~\eqref{eq:PreSubsecscovv} as\footnote{Details are shown in  Appendix~\ref{piapendix}.}
\begin{equation} \label{eq:InvariantProp1}
\tilde{D}^{\mu \nu}_{B}(x,y) = \Omega(x,y) \tilde{D}^{\mu \nu}_{B}(x-y) 
\end{equation}
with
\begin{equation}
    \tilde{D}^{\mu \nu}_{B}(x-y) = \int \frac{d^{4}p}{(2 \pi)^{4}} e^{-ip \cdot (x-y)} \tilde{D}^{\mu \nu}_{B}(p),
\end{equation}
where
\begin{equation} \label{eq:InvariantProp2}
\begin{split}
\tilde{D}^{\mu \nu}_{B}(p) & \equiv - \int^{\infty}_{0} \frac{ds}{\cos{(eBs)}} e^{is \left[ p^{2}_{\parallel} + p^{2}_{\bot} \frac{\tan{(eBs)}}{eBs} - m_{W}^{2} \right]} \\
& \hspace{1cm} \times \Bigg[ \left( - g^{\alpha \beta} \right) \sum_{s_{3}} \tilde{\Delta}^{\mu}{}_{\alpha}(s_{3}) \tilde{\Delta}^{\nu}{}_{\beta}(s_{3}) e^{2i eB s s_{3}} + \frac{1}{m_{W}^{2}} \left( \tilde{p}^{\mu}_{t} \tilde{p}^{\nu \prime}_{t} - \frac{eB}{2} \left[ \tilde{\varphi}^{\mu \nu} - i \tan{(eBs)} g^{\mu \nu}_{\bot} \right] \right) \Bigg].
\end{split}
\end{equation}
In the last equation, we have introduced the notation 
and
\begin{equation}
\begin{split}
p^{2}_{\bot} &\equiv  p^{1} p_{1} + p^{2} p_{2},\\
\tilde{p}^{\mu}_{t} &\equiv \left( p^{0}, p^{-} t_{eB}, p^{+} t_{eB}^{*}, p^{3} \right), 
\\
\tilde{p}^{\mu \prime}_{t} &\equiv \left( p^{0}, p^{+} t_{eB}, p^{-} t_{eB}^{*}, p^{3} \right), \\
p^{\pm} &\equiv \left( p^{1} \pm i p^{2} \right) / \sqrt{2}, \\
t_{eB} &\equiv i \tan{(eBs)} + 1.
\end{split}
\end{equation}
It is clear in Eq.~\eqref{eq:InvariantProp1} that the Schwinger phase $\Omega(x,y)$ is the only source of non-invariance under translations.

To compare with the more conventional expressions in the literature~\cite{HighEnergyNeutrino_erdas, ElectroweakProcesses_kuznetsov}, we rotate back via
\begin{equation} \label{eq:PropWMagDesrotSchwinger}
\begin{split}
 D^{\mu \nu}_{B}(p) &= \left( N^{-1} \right)^{\mu}{}_{\alpha} \tilde{D}^{\alpha \beta}_{B}(p) N_{\beta}{}^{\nu} \\
 &= -\int_{0}^{\infty} \frac{ds}{\cos{(eBs)}} e^{-is \left[ m^{2}_{W} - p_{\parallel}^{2} - p_{\bot}^{2} \frac{\tan{(eBs)}}{eBs} \right]} \Bigg[g^{\mu \nu}_{\parallel} + g^{\mu \nu}_{\bot} \cos{(2eBs)} + \varphi^{\mu \nu} \sin{(2eBs)} \\
& \hspace{2cm} - \frac{1}{m^{2}_{W}} \left[ \left( p^{\mu} + \varphi^{\mu}{}_{\alpha} p^{\alpha} \tan{(eBs)} \right) \left( p^{\nu} + p^{\beta} \varphi_{\beta}{}^{\nu} \tan{(eBs)} \right) - \frac{i eB}{2} \left( \varphi^{\mu \nu} - g^{\mu \nu}_{\bot} \tan{(eBs)} \right) \right] \Bigg]
\end{split}
\end{equation}
Eq.~\eqref{eq:PropWMagDesrotSchwinger} gives the $W^{-}$-boson propagator in a homogeneous magnetic field of arbitrary strength in the Schwinger proper-time representation and constitutes one of the central results of this work. A direct comparison with existing expressions is complicated by differing metric signature conventions~\cite{HighEnergyNeutrino_erdas}. Nevertheless, within the mostly-minus metric signature convention $g^{\mu\nu}=\mathrm{diag}(+,-,-,-)$ adopted here, we find a slight  discrepancy between Eq.~\eqref{eq:PropWMagDesrotSchwinger} and the corresponding result reported in Refs.~\cite{ElectroweakProcesses_kuznetsov, ChargedMassive_iablokov}.

\section{LSZ Reduction Formula for Charged Vector Bosons in a Magnetic Field} \label{sec.LSZMag}

As an application of the Ritus-eigenfunction representation of the charged vector-boson propagator, we derive the LSZ reduction formula in a constant magnetic background. The LSZ procedure provides for the systematic amputation of charged external legs and is, therefore, essential for computing self-energies and polarization tensors. For the sake of completeness, the conventional vacuum derivation is briefly summarized in Appendix~\ref{sec.LSZB=0}, which we closely follow for the $B\neq 0$ case.

\subsection{LSZ Reduction Formula $B \neq 0$}

Following the ideas shown in Appendix~\ref{sec.LSZB=0}, let us start by writing the transition amplitude for a charged vector boson as follows
\begin{equation} \label{eq:AmpProbMag}
\begin{split}
\bra{f} S \ket{i} & = 2 \sqrt{E_{\bar{p}_{i}} E_{\bar{p}_{f}}} \bra{\Omega} \hat{a}_{\bar{p}_{f}, \lambda_{f}}(\infty) \hat{a}_{\bar{p}_{i}, \lambda_{i}}^{\dagger}(- \infty) \ket{\Omega} \\
\end{split}
\end{equation}
where $\ket{\Omega}$ denotes the interacting vacuum in the presence of the background external magnetic field. Note that the creation and annihilation operators, $\hat{a}^{\dagger}_{\bar{p}, \lambda}(t)$ and $\hat{a}_{\bar{p}, \lambda}(t)$, respectively, are explicitly time-dependent and act on the interacting vacuum $\ket{\Omega}$. The time dependency is incorporated as usual, 
\begin{equation}
\hat{a}_{\bar{p}, \lambda}(t) = e^{i \hat{H} (t - t_{0})} \hat{a}_{\bar{p}, \lambda}(t_{0}) e^{-i \hat{H} (t - t_{0})},
\end{equation}
where $\hat{H}$ is the Hamiltonian, with $\hat{a}_{\bar{p}, \lambda}(t)$ the time evolved annihilation operator from  an initial time $t_{0}$.

Now, using the ``non-rotated'' version of the quantum field operators\footnote{The relation between rotated and ``non-rotated''  bases is established in Appendix~\ref{sec.RitusDesRot}.} in a constant magnetic field background given in Eq.~\eqref{eq:CuantizacionMag}, by multiplying it with the factor $(i\partial_t+E_{\bar{p}'})$, and once some integrations by parts are completed and boundary conditions on the asymptotic states are applied, it follows that the annihilation operator, rewritten in terms of the quantum field,  has the form
\begin{equation}\label{eq:aoppfield}
\begin{split}
&\int d^{3}x \; \left( \mathbb{E}^{+ \dagger}_{\ell, 1} \right)^{\! \! \alpha}_{\, \mu} \! (p, x) \, 
\left( i \partial_{t} + E_{\bar{p}^{\prime}} \right) \hat{W}^{(-) \mu}(x) 
= \sqrt{2 E_{\bar{p}^{\prime}}} \sqrt{2 \pi} \sum_{\lambda = 1}^{3} \hat{a}_{\bar{p}^{\prime}, \lambda}(t) \bar{\varepsilon}_{\mathbf{\lambda}}^{(-) \alpha} \! \left( \bar{p}^{\, \prime}, \ell \right) e^{-i E_{\bar{p}^{\prime}} x_{0}},
\end{split}
\end{equation}
where we have used the orthogonality relations 
\begin{equation}
\begin{split}
& \int dx^{1} \; D_{\ell^{\prime} - 1 + \hat{s}_{3}} (\rho_{\pm}) D_{\ell - 1 + \hat{s}_{3}} (\rho_{\pm}) = \frac{\pm 1}{\sqrt{2|eB|}} \delta_{\ell, \ell^{\prime}} \left( \ell - 1 + \hat{s}_{3} \right)! \sqrt{2 \pi}, \\
\end{split}
\end{equation}
together with the change of variables in Eq.~\eqref{eq:CambioVariable}.

Next, taking into account the relations
\begin{equation}
\bar{\varepsilon}^{\mu}_{\mathbf{\lambda}} \left( \bar{p}, \ell \right) \bar{\varepsilon}_{\mu, \mathbf{\lambda}^{\prime}}^{*} \left( \bar{p}, \ell \right) = - \delta_{\lambda, \lambda^{\prime}}, \hspace{4cm} \bar{\varepsilon}^{\mu}_{\mathbf{\lambda}} \left(\bar{p}, \ell \right) \bar{\varepsilon}_{\mu, \mathbf{\lambda}^{\prime}} \left( \bar{p}, \ell \right) = 0,
\end{equation}
then, it follows that once we multiply Eq.~\eqref{eq:aoppfield} by the vector polarization $\bar{\varepsilon}^{(-)*}_{\alpha, \mathbf{\lambda}^{\prime}} \! \left( \bar{p}, \ell \right)$, we obtain
\begin{equation}
\begin{split}
& \hat{a}_{\bar{p}, \lambda}(t) = \frac{-1}{\sqrt{2E_{\bar{p}}} \sqrt{2 \pi}} \int d^{3}x \; \left( \mathbb{E}^{+ \dagger}_{\ell, 1} \right)^{\! \! \alpha}_{\, \mu} \! (p,x) \, \bar{\varepsilon}^{(-) *}_{\alpha, \mathbf{\lambda}} \! \left( \bar{p}, \ell \right) \left( i \partial_{t} + E_{\bar{p}} \right) \hat{W}^{(-) \mu}(x), \\
\end{split}
\end{equation}
where the ``non-rotated" Ritus eigenfunction is defined in Eq. (\ref{eq:EigenFunDesRot}).  Proceeding as in the vacuum case, 
now, by taking into account the identity
\begin{equation}\label{ideinttB}
\sqrt{2E_{\bar{p}}} \left[ \hat{a}_{\bar{p},\lambda} (\infty) - \hat{a}_{\bar{p},\lambda}(-\infty) \right] = \int_{-\infty}^{\infty} dt \;  \partial_{t} a_{\bar{p},\lambda}(t),
\end{equation}
then, the magnetic field equivalent to Eq.~\eqref{eq:DifOpCreaAniqVac}, yields 
\begin{equation} \label{eq:DifOpCreaAniqMag}
\begin{split}
\sqrt{2E_{\bar{p}}} \left[ \hat{a}_{\bar{p}, \lambda}(\infty) - \hat{a}_{\bar{p}, \lambda}(-\infty) \right] & = \frac{-i}{\sqrt{2 \pi}} \int d^{4}x \; \left( \mathbb{E}^{+ \dagger}_{\ell, 1} \right)^{\! \! \alpha}_{\,\mu} \! (p,x) \, \bar{\varepsilon}^{(-) *}_{\alpha, \mathbf{\lambda}} \! \left( \bar{p}, \ell \right) \left[ \partial_{t}^{2} + E_{\bar{p}}^{2} \right] \hat{W}^{(-) \mu}(x). \\
\end{split}
\end{equation}
At this point, the appearance of $\partial_{t}^{2}+E_{p}^{2}$ is natural from the projection onto $\mathbb{E}_{\ell}$ waves. 

Next, to relate the last equation with EOM, we perform a Fourier analysis on Eq.~\eqref{eq:EOM1}, and multiplying it by $\bar{\varepsilon}^{(-) *}_{\mu, \mathbf{\lambda}} \! (\bar{p}, \ell)$, getting
\begin{equation}\label{eq:DifOpCreaAniqMag2s}
\begin{split}
& \int d^{4}x \; \left( \mathbb{E}^{+ \dagger}_{\ell, 1} \right)^{\! \! \mu}_{\, \alpha} \! (p, x) \, \bar{\varepsilon}^{(-) *}_{\mu, \mathbf{\lambda}} \! \left( \bar{p}, \ell \right) \sum_{s_{3}^{\prime}} \left( D^{\sigma} D_{\sigma} + m_{W}^{2} - 2eB s_{3}^{\prime} \right) \Delta^{\alpha}{}_{\beta}(s_{3}^{\prime}) \hat{W}^{(-) \beta}(x) \\
&\hspace{5cm}= \int d^{4}x \; \left( \mathbb{E}^{+ \dagger}_{\ell, 1} \right)^{\! \! \mu}_{\, \beta} \! (p,x) \, \bar{\varepsilon}^{(-) *}_{\mu, \mathbf{\lambda}} \! \left( \bar{p}, \ell \right) \left[ \partial^{2}_{t} + E^{2}_{\bar{p}} \right] \hat{W}^{(-) \beta}(x).
\end{split}
\end{equation}
where two integrations by parts over the space were performed to get the {\it rhs} of last equation. 

Comparing Eq.~\eqref{eq:DifOpCreaAniqMag} and Eq.~\eqref{eq:DifOpCreaAniqMag2s}, we arrive at 
\begin{equation}\label{eq:beforelsz}
\begin{split}
\int d^{4}x \; \left( \mathbb{E}^{+ \dagger}_{\ell, 1} \right)^{\! \! \mu}_{\, \alpha} \! (p, x) \, \bar{\varepsilon}^{(-) *}_{\mu, \mathbf{\lambda}} \! \left( \bar{p}, \ell \right) \sum_{s_{3}^{\prime}} \left( D^{\sigma} D_{\sigma} + m_{W}^{2} - 2eB s_{3}^{\prime} \right) \Delta^{\alpha}{}_{\beta}(s_{3}^{\prime}) \hat{W}^{(-) \beta}(x)& \\
&\hspace{-5em}=i \sqrt{2 \pi} \sqrt{2E_{\bar{p}}} \left[ \hat{a}_{\bar{p}, \lambda}(\infty) - \hat{a}_{\bar{p}, \lambda}(- \infty) \right].
\end{split}
\end{equation}

Substituting Eq.~\eqref{eq:beforelsz} together with its adjoint into Eq.~\eqref{eq:AmpProbMag}, one obtains the LSZ reduction formula in a magnetic background,
\begin{equation} \label{eq:LSZFinal}
\begin{split}
\bra{f} S \ket{i}& = \frac{1}{2 \pi} \int d^{4} x_{f} \; \left( \mathbb{E}_{\ell, 1}^{+ \dagger} \right)^{\! \! \mu}_{\, \alpha} \! (p_{f}, x_{f}) \,  \bar{\varepsilon}^{(-) *}_{\mu, \mathbf{\lambda_{f}}} \left( \bar{p}_{f}, \ell_{f} \right) \left( \bar{p}_{f}^{\sigma} \bar{p}^{*}_{f, \sigma} - m_{W}^{2} \right) \\
& \times \int d^{4} x_{i} \;  \left( \mathbb{E}_{\ell, 1}^{+} \right)^{\! \! \nu}_{\, \beta} \! (p_{i}, x_{i}) \, \bar{\varepsilon}^{(-)}_{\nu, \mathbf{\lambda_{i}}} \left( \bar{p}_{i}, \ell_{i} \right) \left( \bar{p}^{\rho}_{i} \bar{p}_{\rho,i}^{*} - m_{W}^{2} \right) \bra{\Omega} \mathcal{T} \left\{ \hat{W}^{(-) \alpha}(x_{f}) \hat{W}^{(+) \beta}(x_{i}) \right\} \ket{\Omega}.
\end{split}
\end{equation}
This equation, which is a novel result, provides the explicit LSZ amputation procedure for charged massive vector bosons in a constant magnetic field, expressed directly in the Ritus basis. The result supplies the missing ingredient required to relate polarization tensors computed in a magnetic background to physical on-shell amplitudes for charged vector bosons. For instance, in Refs.~\cite{EffectiveMass_skalozub, ChargedAndNeutral_liu} the charged vector-boson polarization tensor is evaluated in a constant magnetic field, but the external legs are not explicitly amputated. Eq.~\eqref{eq:LSZFinal} provides the appropriate LSZ reduction step needed to complete such calculations at the level of S-matrix elements.

\section{Conclusions}
\label{sec.concl}

In this work, we have derived the propagator of a massive charged vector boson in the presence of a constant magnetic field of arbitrary strength, working in the mostly-minus metric and adopting the unitary gauge. To our knowledge, this is the first time of the charged vector-boson propagator in the unitary gauge within the Ritus eigenfunction framework, rendering it particularly suitable for contemporary applications such as neutrino self-energy calculations in strong magnetic backgrounds.

A central technical ingredient in the presented construction is the explicit determination of a consistent polarization basis in the magnetic background. The polarization vectors depend non-trivially on the Landau level, reflecting the reduction of available spin projection configurations at low excitation and allowing one or two additional physical spin projection modes at higher levels. Since different choices of basis can obscure physical interpretation or complicate practical calculations, we provided the polarization sets explicitly in the main text and presented their derivation in detail in Appendix~\ref{sec.PolVec}. This last is essential for reproducibility and for subsequent applications that require controlled polarization sums and external state projections.

In addition, we derived the LSZ reduction formula for a charged massive vector boson in a constant magnetic field. This provides the amputation procedure for external charged vector states dressed by the background field, and is therefore essential in the calculation of amplitudes and for defining quantities associated with quantum fluctuations, such as self-energies and vertex corrections. The derivation exploits the Ritus basis to parallel the vacuum construction as closely as possible, with plane-wave external modes replaced by the corresponding Ritus eigenfunctions. This result supplies a missing step in several existing analyses of charged vector polarization tensors in magnetic backgrounds, where the amputated amplitudes were not explicitly constructed.

We also presented a systematic method to connect the Ritus representation to the Schwinger proper-time representation. The procedure proceeds by introducing the proper-time parametrization, summing the Ritus expression over all Landau levels, isolating the Schwinger phase that accounts for the residual gauge dependence in coordinate space, and identifying the remaining transverse structures as Gaussian integrals. This sequence leads to the proper-time form in Eq.~\eqref{eq:PropWMagDesrotSchwinger}. Within this mapping, we found a slight  discrepancy with the corresponding expression reported in Ref.~\cite{ElectroweakProcesses_kuznetsov} when working in the mostly-minus signature. The difference resides in terms associated with the unitary-gauge structure of the propagator; in contrast, no mismatch arises in the Feynman gauge. The origin and implications of this slight discrepancy merit further investigation, particularly in computations of self-energies and related observables for particles coupled to charged vector bosons under the influence of a magnetic field.

Several extensions of the present work are natural. An immediate application is a recalculation of the neutrino self-energy in a magnetic field within the unitary gauge, using the propagator derived here and assessing quantitatively the impact of the slight discrepancy identified relative to existing proper-time expressions. Such studies would further clarify the gauge dependence of electroweak processes under external magnetic field.

\acknowledgements

\noindent Support for this work has been received from SECIHTI fellowship No. 1184435 and UNAM-PAPIIT under
Grant No. IN108123.


\appendix

\section{Symmetry Transformations} \label{sec.Symmetry}

When dealing with charged particles it is important to specify conventions and transformation properties with care, in order to avoid ambiguities arising from different sign or phase conventions. Starting from the Lagrangian in Eq.~\eqref{eq:Lagrangian}, it is important to note that the residual custodial $U(1)$ symmetry is preserved. This symmetry implies the local gauge transformations
\begin{equation}
\begin{split}
& \psi(x) \rightarrow \psi'(x) = e^{i q \theta(x)} \psi(x), \\
& \bar{\psi}(x) \rightarrow \bar{\psi}{}^{'}(x) = e^{-i q \theta(x)} \bar{\psi}(x), \\
& A^{\mu}(x) \rightarrow A^{\mu \prime} = A(x) - \frac{q}{e} \partial^{\mu} \theta(x)
\end{split}
\end{equation}
where $\psi$ denotes a charged spinor field, $\theta(x)$ is an arbitrary local scalar function, and $q$ is the parameter associated with the $U(1)$ transformation. With this, the covariant derivative is defined as
\begin{equation} \label{eq:CovDev}
D^{\mu} \equiv \partial^{\mu} + i e A^{\mu}(x).
\end{equation}
where the electric charge $e$ may be positive or negative, depending on the particle under consideration. Under a local $U(1)$ transformation one finds
\begin{equation}
\left( D^{\mu} \psi(x) \right)' = D^{\mu '} \psi'(x) = D^{\mu '} e^{i q \theta(x)} \psi(x).
\end{equation}

Gauge covariance requires that the covariant derivative transforms in the same way as the field itself~\cite{GaugeTheory_cheng}, namely
\begin{equation}
\begin{split}
& \left( D^{\mu} \psi(x) \right)' = e^{i q \theta(x)} D^{\mu} \psi(x), \\
& D^{\mu '} = e^{i q \theta(x)} D^{\mu} e^{- i q \theta(x)} \\
&\hspace{2em}= \partial^{\mu} - i q \partial^{\mu} \theta(x) + i e A^{\mu} (x).
\end{split}
\end{equation}

After applying the transformation of the gauge field,
\begin{equation}  \label{eq:transfAmu}
\begin{split}
& A^{\mu}(x) \rightarrow A^{\mu '} = A^{\mu}(x) - \frac{q}{e} \partial^{\mu} \theta(x),
\end{split}
\end{equation}
the covariant derivative assumes the expected form
\begin{equation}  \label{eq:transfAmu2}
\begin{split}
D^{\mu '} = \partial^{\mu} + i e A^{\mu '}(x).
\end{split}
\end{equation}
With these definitions and transformation properties explicitly stated, potential inconsistencies in the treatment of charged fields and their couplings to the electromagnetic background would, hopefully, be avoided.

\section{Calculations for the polarization vectors} \label{sec.PolVec}

Obtaining the polarization vectors associated with different energy eigenstates, and therefore with different Landau levels, is a central ingredient of the present work. For this reason, the procedure leading to the explicit construction of the polarization vector sets is presented here in full detail, as outlined in Sec.~\ref{sec.WMF}, and organized according to the value of the Landau level $\ell$. The analysis carried out in this appendix is restricted to the polarization vectors $\tilde{\bar{\varepsilon}}^{(-) \mu}_{\mathbf{\lambda}}$ of the negatively charged vector boson. The corresponding vectors for the positively charged field follow straightforwardly from complex conjugation, $\tilde{\bar{\varepsilon}}^{(-) \mu*}_{\mathbf{\lambda}}=\tilde{\bar{\varepsilon}}^{(+) \mu}_{\mathbf{\lambda}}$. Moreover, when reverting to the ``non--rotated'' representation it is useful to recall that  $ \tilde{\bar{\varepsilon}}^{(-) \mu}_{\mathbf{\lambda}} \propto \tilde{W}^{(-) \mu}(x) \equiv N^{\mu}{}_{\alpha}W^{(-) \alpha}(x) \propto N^{\mu}{}_{\alpha} \bar{\varepsilon}^{(-) \alpha}_{\mathbf{\lambda}}$. Also, for simplicity sake, the momentum and Landau-level dependence of the polarization vectors will be omitted.

\noindent Writing Eq. (\ref{eq:transv}) explicitly for $r = +1$, yields

\begin{equation}
\begin{split}
X^{+}_{\ell - 1}(\rho_{+}) e^{-i \left( E x_{0} + p^{2} x_{2} + p^{3} x_{3} \right)} \left[ E \tilde{\bar{\varepsilon}}_{0, \mathbf{\lambda}}^{(-)} + i s_{eB} \sqrt{|eB|} \sqrt{\ell_{+}} \tilde{\bar{\varepsilon}}^{(-)}_{1, \mathbf{\lambda}} - i s_{eB} \sqrt{|eB|} \sqrt{\ell_{-}} \tilde{\bar{\varepsilon}}_{2, \mathbf{\lambda}}^{(-)} + p^{3} \tilde{\bar{\varepsilon}}_{3, \mathbf{\lambda}}^{(-)} \right] = 0.
\end{split}
\end{equation}

For $\ell\geq1$ the parabolic cylinder functions satisfy
$X^{+}_{\ell - 1}(\rho_{+}) \neq 0$, and therefore the transversality condition reduces to
\begin{equation} \label{eq:GaugeCondExpanded}
E \tilde{\bar{\varepsilon}}_{0, \mathbf{\lambda}}^{(-)} + i s_{eB} \sqrt{|eB|} \sqrt{\ell_{+}} \tilde{\bar{\varepsilon}}^{(-)}_{1, \mathbf{\lambda}} - i s_{eB} \sqrt{|eB|} \sqrt{\ell_{-}} \tilde{\bar{\varepsilon}}_{2, \mathbf{\lambda}}^{(-)} + p^{3} \tilde{\bar{\varepsilon}}_{3, \mathbf{\lambda}}^{(-)} = 0.
\end{equation}

A second independent condition follows from the orthonormalization relation,
Eq.~\eqref{def:ortonor}, which can be written explicitly as
\begin{equation} \label{eq:OrthonormalizationExpanded}
\begin{split}
& \int d\rho_{r} \; 2E \sqrt{2 \pi} \Bigg[ \frac{\left( D_{\ell - 1}(\rho_{+}) \right)^{2}}{\left( \ell - 1 \right)!} \left( \tilde{\bar{\varepsilon}}^{(-) 0*}_{\mathbf{\lambda}^{\prime}} \, \tilde{\bar{\varepsilon}}^{(-) 0}_{\mathbf{\lambda}} - \tilde{\bar{\varepsilon}}^{(-) 3*}_{\mathbf{\lambda}^{\prime}} \, \tilde{\bar{\varepsilon}}^{(-) 3}_{\mathbf{\lambda}} \right) \\
& \hspace{3cm} - \frac{\left( D_{\ell - 1 + s_{eB}}(\rho_{+}) \right)^{2}}{\left( \ell - 1 + s_{eB} \right)!} \tilde{\bar{\varepsilon}}^{(-) 1*}_{\mathbf{\lambda}^{\prime}} \, \tilde{\bar{\varepsilon}}^{(-) 1}_{\mathbf{\lambda}} - \frac{\left( D_{\ell - 1 - s_{eB}}(\rho_{+}) \right)^{2}}{\left( \ell - 1 - s_{eB} \right)!} \tilde{\bar{\varepsilon}}^{(-) 2*}_{\mathbf{\lambda}^{\prime}} \, \tilde{\bar{\varepsilon}}^{(-) 2}_{\mathbf{\lambda}} \Bigg] = - 4\pi E \delta_{\mathbf{\lambda}, \mathbf{\lambda}^{\prime}}.
\end{split}
\end{equation}
This last equation holds for all values of $\ell$. In the following, the cases
$\ell=0$, $\ell=1$, and $\ell>1$ are discussed separately for $s_{eB}=+1$.
The case $s_{eB}=-1$ follows analogously, and the general expressions are obtained
straightforwardly from these results.

\subsection{$\ell = 0$ case}

For $\ell=0$, Eq.~\eqref{eq:OrthonormalizationExpanded} reduces to
\begin{equation} \label{eq:Orthol0}
\begin{split}
\int d\rho_{r} \; 2E \sqrt{2 \pi} \left[ \frac{\left( D_{0}(\rho_{+}) \right)^{2}}{\left( 0 \right)!} \tilde{\bar{\varepsilon}}^{(-) 1*}_{\mathbf{\lambda}^{\prime}} \, \tilde{\bar{\varepsilon}}^{(-) 1}_{\mathbf{\lambda}} \right] = 4\pi E \delta_{\mathbf{\lambda}, \mathbf{\lambda}^{\prime}} 
\end{split}
\end{equation}
which implies $\tilde{\bar{\varepsilon}}^{(-) 1*}_{\mathbf{\lambda}^{\prime}} \, \tilde{\bar{\varepsilon}}^{(-) 1}_{\mathbf{\lambda}} = 1$ for $\lambda=\lambda^{\prime}$. No additional constraints arise, and therefore only a single polarization
vector exists at this level,
\begin{equation}
\tilde{\bar{\varepsilon}}^{(-) \mu}_{\mathbf{1}} \equiv \left( 0, s_{+}, s_{-}, 0 \right),
\end{equation}
where the polarization index $\mathbf{\lambda} = \mathbf{1}$ emphasizes that this is the only polarization vector possible in this energy state.

\subsection{$\ell = 1$ case}

For $\ell=1$, the transversality condition, Eq. (\ref{eq:GaugeCondExpanded}), becomes
\begin{equation} \label{eq:Transversalityl1}
E \tilde{\bar{\varepsilon}}_{0, \mathbf{\lambda}}^{(-)} + i \sqrt{|eB|}  \tilde{\bar{\varepsilon}}^{(-)}_{1, \mathbf{\lambda}} - i \sqrt{|eB|} \tilde{\bar{\varepsilon}}^{(-)}_{2, \mathbf{\lambda}} + p^{3} \tilde{\bar{\varepsilon}}^{(-)}_{3, \mathbf{\lambda}} = 0,
\end{equation}
while the orthonormalization condition, Eq. (\ref{eq:OrthonormalizationExpanded}), yields
\begin{equation} \label{eq:GaugeCondl1}
\begin{split}
& \int d\rho_{r} \; 2E \sqrt{2 \pi} \left[ \frac{\left( D_{0}(\rho_{+}) \right)^{2}}{\left( 0 \right)!} \left( \tilde{\bar{\varepsilon}}^{(-) 0*}_{\mathbf{\lambda}^{\prime}} \, \tilde{\bar{\varepsilon}}^{(-) 0}_{\mathbf{\lambda}} - \tilde{\bar{\varepsilon}}^{(-) 3*}_{\mathbf{\lambda}^{\prime}} \, \tilde{\bar{\varepsilon}}^{(-) 3}_{\mathbf{\lambda}} \right) - \frac{\left( D_{1}(\rho_{+}) \right)^{2}}{\left( 1 \right)!} \tilde{\bar{\varepsilon}}^{(-) 1*}_{\mathbf{\lambda}^{\prime}} \,  \tilde{\bar{\varepsilon}}^{(-) 1}_{\mathbf{\lambda}} \right] = - 4\pi E \delta_{\mathbf{\lambda}, \mathbf{\lambda}^{\prime}} \\
& \hspace{3cm} \implies \tilde{\bar{\varepsilon}}^{(-) 0 *}_{\mathbf{\lambda}^{\prime}} \, \tilde{\bar{\varepsilon}}^{(-) 0}_{\mathbf{\lambda}} - \tilde{\bar{\varepsilon}}^{(-) 1 *}_{\mathbf{\lambda}^{\prime}} \, \tilde{\bar{\varepsilon}}^{(-) 1}_{\mathbf{\lambda}} - \tilde{\bar{\varepsilon}}^{(-) 3 *}_{\mathbf{\lambda}^{\prime}} \, \tilde{\bar{\varepsilon}}^{(-) 3}_{\mathbf{\lambda}} = - \delta_{\mathbf{\lambda}, \mathbf{\lambda}^{\prime}}
\end{split}
\end{equation}

A convenient ansatz for the first polarization vector is
\begin{equation}
\tilde{\bar{\varepsilon}}^{(-) \mu}_{\mathbf{1}} = \alpha \left( p^{3}, 0, 0, E \right),
\end{equation}
so that after taking $\mathbf{\lambda} = \mathbf{1} = \mathbf{\lambda}^{\prime}$
\begin{equation}
\tilde{\bar{\varepsilon}}^{(-) \mu}_{\mathbf{1}}= \sqrt{\frac{1}{m_{\bot}^{2}}} \left( p^{3}, 0, 0, E \right),
\end{equation}
where $m_{\bot}^{2} \equiv E^{2} - (p^{3})^{2}$. Considering the most general polarization vector for the second one
\begin{equation}
\tilde{\bar{\varepsilon}}^{(-) \mu}_{\mathbf{2}} = \beta \left( \tilde{\bar{\varepsilon}}^{(-) 0}_{\mathbf{2}} , \tilde{\bar{\varepsilon}}^{(-) 1}_{\mathbf{2}}, \tilde{\bar{\varepsilon}}^{(-) 2}_{\mathbf{2}}, \tilde{\bar{\varepsilon}}^{(-) 3}_{\mathbf{2}} \right),
\end{equation}
and after using the transversality condition, Eq. (\ref{eq:Transversalityl1}), and orthonormalization condition, Eq. (\ref{eq:GaugeCondl1}), we get
\begin{equation}
\tilde{\bar{\varepsilon}}^{(-) \mu}_{\mathbf{2}} = \sqrt{\frac{|eB|}{m_{\bot}^{2} \left( m_{\bot}^{2} - |eB| \right)}} \left(E, \frac{-i m_{\bot}^{2}}{\sqrt{|eB|}}, 0, p^{3} \right),
\end{equation}
where no conditions exist over the entry $\tilde{\bar{\varepsilon}}^{(-) 2}_{\mathbf{2}}$ so it can be taken null for simplicity.

\noindent Analogously, for $s_{eB} = -1$

\begin{equation}
\tilde{\bar{\varepsilon}}^{(-) \mu}_{\mathbf{2}} = \sqrt{\frac{|eB|}{m_{\bot}^{2} \left( m_{\bot}^{2} - |eB| \right)}} \left(E, 0, \frac{-i m_{\bot}^{2}}{\sqrt{|eB|}}, p^{3} \right).
\end{equation}

\noindent For this reason, the polarization vectors that appear in the $\ell = 1$ Landau level are

\begin{equation}
\begin{split}
& \tilde{\bar{\varepsilon}}^{(-) \mu}_{\mathbf{1}} = \sqrt{\frac{1}{m_{\bot}^{2}}} \left( p^{3}, 0, 0, E \right), \\
& \tilde{\bar{\varepsilon}}^{(-) \mu}_{\mathbf{2}} = \sqrt{\frac{|eB|}{m_{\bot}^{2} \left( m_{\bot}^{2} - |eB| \right)}} \left(E, \frac{-i m_{\bot}^{2}}{\sqrt{|eB|}} s_{+}, \frac{-i m_{\bot}^{2}}{\sqrt{|eB|}} s_{-}, p^{3} \right).
\end{split}
\end{equation}

\subsection{$\ell > 1$ case}

For higher Landau levels, three independent polarization vectors exist, as expected for a massive spin-1 particle. The first polarization can again be chosen as
\begin{equation}
\tilde{\bar{\varepsilon}}^{(-) \mu}_{\mathbf{1}} = \sqrt{\frac{1}{m_{\bot}^{2}}} \left( p^{3}, 0, 0, E \right).
\end{equation}

The remaining two polarizations are taken in the generic form
\begin{equation}
\begin{split}
& \tilde{\bar{\varepsilon}}^{(-) \mu}_{\mathbf{2}} = \beta \left( \tilde{\bar{\varepsilon}}^{(-) 0}_{\mathbf{2}} , \tilde{\bar{\varepsilon}}^{(-) 1}_{\mathbf{2}}, \tilde{\bar{\varepsilon}}^{(-) 2}_{\mathbf{2}}, \tilde{\bar{\varepsilon}}^{(-) 3}_{\mathbf{2}} \right), \\
& \tilde{\bar{\varepsilon}}^{(-) \mu}_{\mathbf{3}} = \gamma \left( \tilde{\bar{\varepsilon}}^{(-) 0}_{\mathbf{3}} , \tilde{\bar{\varepsilon}}^{(-) 1}_{\mathbf{3}}, \tilde{\bar{\varepsilon}}^{(-) 2}_{\mathbf{3}}, \tilde{\bar{\varepsilon}}^{(-) 3}_{\mathbf{3}} \right),
\end{split}
\end{equation}
and are fixed by imposing: (i) orthogonality to $\tilde{\bar{\varepsilon}}^{(-)}_{\mathbf{1}}$, (ii) transversality \eqref{eq:GaugeCondExpanded}, (iii) mutual orthogonality, and (iv) normalization. From orthogonality with $\tilde{\bar{\varepsilon}}^{(-)}_{\mathbf{1}}$, we obtain
\begin{equation}
\begin{split}
& \tilde{\bar{\varepsilon}}^{(-) 0}_{\mathbf{2}} = \frac{E}{p^{3}} \tilde{\bar{\varepsilon}}^{(-) 3}_{\mathbf{2}}, \\
& \tilde{\bar{\varepsilon}}^{(-) 0}_{\mathbf{3}} = \frac{E}{p^{3}} \tilde{\bar{\varepsilon}}^{(-) 3}_{\mathbf{3}}.
\end{split}
\end{equation}
respectively. On the other hand, the transversality condition, Eq. (\ref{eq:GaugeCondExpanded}), yields 
\begin{equation}
\tilde{\bar{\varepsilon}}^{(-) 1}_{\mathbf{2}} = \frac{-i m_{\bot}^{2}}{p^{3} \sqrt{|eB| \ell}} \tilde{\bar{\varepsilon}}^{(-) 3}_{\mathbf{2}},
\end{equation}
where $\tilde{\bar{\varepsilon}}^{(-) 2}_{\mathbf{2}} = 0$ is chosen for simplicity, since no conditions exist over this entry. Finally,  using normalization, we fixed the overall scale, getting
\begin{equation}
\tilde{\bar{\varepsilon}}^{(-) \mu}_{\mathbf{2}} = \sqrt{\frac{1}{m_{\bot}^{2}}} \sqrt{\frac{|eB| \ell}{m_{\bot}^{2} - |eB| \ell}} \left( E, \frac{-i m_{\bot}^{2}}{\sqrt{|eB| \ell}}, 0, p^{3} \right).
\end{equation}

For the third polarization, the transversality condition, Eq. (\ref{eq:GaugeCondExpanded}), gives
\begin{equation}
\tilde{\bar{\varepsilon}}^{(-) 1}_{\mathbf{3}} = \frac{-i \sqrt{|eB| \ell}}{p^{3}} \tilde{\bar{\varepsilon}}^{(-) 3}_{\mathbf{3}}.
\end{equation}
and the remaining transverse component is fixed by imposing orthogonality to $\tilde{\bar{\varepsilon}}^{(-)}_{\mathbf{2}}$ together with normalization. This leads to
\begin{equation}
\tilde{\bar{\varepsilon}}^{(-) \mu}_{\mathbf{3}} = \sqrt{\frac{1}{|eB| \ell - m_{\bot}^{2} + \frac{\left( |eB| \ell - m_{\bot}^{2} \right)^{2}}{|eB| (\ell - 1)}}} \left( E, -i \sqrt{|eB| \ell}, -i \frac{|eB| \ell - m_{\bot}^{2}}{\sqrt{|eB| (\ell - 1)}}, p^{3} \right).
\end{equation}

\noindent A similar procedure can be done for $s_{eB} = -1$; thus, the general case for all the polarization vectors in the $\ell > 1$ case is
\begin{equation}
\begin{split}
& \tilde{\bar{\varepsilon}}^{(-) \mu}_{\mathbf{1}} = \sqrt{\frac{1}{m_{\bot}^{2}}} \left( p^{3}, 0, 0, E \right), \\
& \tilde{\bar{\varepsilon}}^{(-) \mu}_{\mathbf{2}} = \sqrt{\frac{1}{m_{\bot}^{2}}} \sqrt{\frac{|eB| \ell}{m_{\bot}^{2} - |eB| \ell}} \left( E, \frac{-i m_{\bot}^{2}}{\sqrt{|eB| \ell}} s_{+}, \frac{-i m_{\bot}^{2}}{\sqrt{|eB| \ell}} s_{-}, p^{3} \right), \\
& \tilde{\bar{\varepsilon}}^{(-) \mu}_{\mathbf{3}} = \sqrt{\frac{1}{|eB| \ell - m_{\bot}^{2} + \frac{\left( |eB| \ell - m_{\bot}^{2} \right)^{2}}{|eB| (\ell - 1)}}} \Bigg( E, -i \sqrt{|eB| \ell} s_{+} - i \frac{|eB| \ell - m_{\bot}^{2}}{\sqrt{|eB| (\ell - 1)}} s_{-},  -i \sqrt{|eB| \ell} s_{-} - i \frac{|eB| \ell - m_{\bot}^{2}}{\sqrt{|eB| (\ell - 1)}} s_{+}, p^{3} \Bigg).
\end{split}
\end{equation}
It is worth noting that the number of admissible polarization vectors depends on the Landau level: for $\ell=0$ only one polarization exists, for $\ell=1$ two polarizations exist, and for $\ell>1$ the full set of three polarizations is recovered. This pattern is a direct manifestation of the magnetic-field-induced splitting of spin states (Zeeman effect).

\section{LSZ Reduction Formula $B = 0$} \label{sec.LSZB=0}

In order to motivate and systematize the derivation of the LSZ reduction formula in a magnetic background, we briefly recount the standard vacuum construction for a charged massive vector boson.

Following the conventional operator approach, let us start by writing  the matrix $S$ for a one particle in to one particle out process involving a charged massive vector boson
in terms of asymptotic creation and annihilation operators as~\cite{QuantumFieldTheory_itzykson}
\begin{equation}\label{eq:AmpProb}
\begin{split}
\bra{f} S \ket{i} & = 2 \sqrt{E_{p_{i}} E_{p_{f}}} \bra{\Omega} \hat{a}_{p_{f}, \lambda_{f}}(\infty) \hat{a}_{p_{i}, \lambda_{i}}^{\dagger}(- \infty) \ket{\Omega} \\
& = 2 \sqrt{E_{p_{i}} E_{p_{f}}} \bra{\Omega} \mathcal{T} \Big \{ \left[ \hat{a}_{p_{f}, \lambda_{f}}(\infty) - \hat{a}_{p_{f}, \lambda_{f}}(-\infty) \right] \left[ \hat{a}_{p_{i}, \lambda_{i}}^{\dagger}(\infty) - \hat{a}_{p_{i}, \lambda_{i}}^{\dagger}(-\infty) \right] \Big \} \ket{\Omega}.
\end{split}
\end{equation}
where the second line follows from inserting time ordered differences of asymptotic operators, which is the standard starting point for obtaining the reduction formula. Note that the creation and annihilation operators, $\hat{a}^{\dagger}_{p, \lambda}(t)$ and $\hat{a}_{p, \lambda}(t)$, respectively, are explicitly time-dependent and act on the interacting vacuum $\ket{\Omega}$. The time dependency is incorporated as usual, 
\begin{equation}
\hat{a}_{p, \lambda}(t) = e^{i \hat{H} (t - t_{0})} \hat{a}_{p, \lambda}(t_{0}) e^{-i \hat{H} (t - t_{0})},
\end{equation}
where $\hat{H}$ is the Hamiltonian, with $\hat{a}_{p, \lambda}(t)$ the time evolved annihilation operator from  an initial time $t_{0}$.

In vacuum, the quantum field operators admit the plane wave expansion
\begin{equation}
\begin{split}
& \hat{W}^{(-) \mu}(x) = \int \frac{d^{3}p}{(2 \pi)^{3}} \frac{1}{\sqrt{2 E_{p}}} \sum_{\lambda = 1}^{3} \left[ \hat{a}_{p,\lambda}(t) \varepsilon^{(-) \mu}_{\lambda} (p) e^{-i p \cdot x} + \hat{b}_{p, \lambda}^{\dagger}(t) \varepsilon^{(+) \mu *}_{\lambda} (p) e^{i p \cdot x} \right], \\
& \hat{W}^{(+) \mu}(x) = \int \frac{d^{3}p}{(2 \pi)^{3}} \frac{1}{\sqrt{2 E_{p}}} \sum_{\lambda = 1}^{3} \left[ \hat{a}^{\dagger}_{p,\lambda}(t) \varepsilon^{(-) \mu *}_{\lambda}(p) e^{i p \cdot x} + \hat{b}_{p, \lambda}(t) \varepsilon^{(+) \mu}_{\lambda} (p) e^{-i p \cdot x} \right],
\end{split}
\end{equation}
where $E_{p}=\sqrt{\vec{p}^{\,2}+M^{2}}$ and $\lambda$ labels the three physical polarizations of a massive spin-1 field. The polarization vectors satisfy
\begin{equation}
    \varepsilon^{(\pm) \mu *}_{\lambda}(p) \varepsilon_{\mu, \lambda^{\prime}}^{(\pm)}(p) = - \delta_{\lambda, \lambda^{\prime}}, \hspace{4cm} \varepsilon^{(\pm) \mu}_{\lambda}(p) \varepsilon_{\mu, \lambda^{\prime}}^{(\pm)}(p) = 0,
\end{equation}
which ensures completeness in the physical subspace and fixes our normalization conventions.

To isolate the annihilation particle operators,  $\hat{a}_{p,\lambda}$, we perform the standard Fourier analysis. Acting with $(i\partial_{t}+E_{p})$ on $\hat{W}^{(-)\mu}$ and integrating against $e^{-i\vec{p}\cdot\vec{x}}$ in such a way that it projects onto the positive-frequency mode,
\begin{equation}
\begin{split}
\int d^{3}x \; e^{-i \vec{p} \cdot \vec{x}} \left( i \partial_{t} + E_{p} \right) \hat{W}^{(-) \mu}(x) & = \sqrt{2 E_{p}} \sum_{\lambda = 1}^{3} \left[ \hat{a}_{p, \lambda}(t) \varepsilon^{(-) \mu}_{\lambda}(p) e^{-i E_{p}t} \right],
\end{split}
\end{equation}

Contracting with $\varepsilon^{(-)*}_{\mu, \lambda}(p)$ and using orthonormality yields
\begin{equation}\label{aaat}
\begin{split}
\hat{a}_{p, \lambda}(t) & = \frac{-1}{\sqrt{2E_{p}}} \int d^{3}x \; e^{i p \cdot x} \varepsilon^{(-) *}_{\mu, \lambda}(p) \left( i \partial_{t} + E_{p} \right) \hat{W}^{(-) \mu}(x).
\end{split}
\end{equation}
In a similar fashion, the creation particle operator can be isolated, yielding
\begin{equation}
\begin{split}
\hat{a}^{\dagger}_{p, \lambda}(t) & = \frac{-1}{\sqrt{2E_{p}}} \int d^{3}x \; e^{-i p \cdot x} \varepsilon_{\mu, \lambda}^{(-)}(p) \left(- i \partial_{t} + E_{p} \right) \hat{W}^{(+) \mu}(x).
\end{split}
\end{equation}

Now, by taking into account the identity
\begin{equation}\label{ideintt}
\sqrt{2E_{p}} \left[ \hat{a}_{p, \lambda} (\infty) - \hat{a}_{p, \lambda}(-\infty) \right] = \int_{-\infty}^{\infty} dt \;  \partial_{t} \hat{a}_{p, \lambda}(t)
\end{equation}
then, using Eq.~\eqref{aaat} on the {\it rhs} of Eq.~\eqref{ideintt} and integrating by parts in time in the usual way, we obtain
\begin{equation} \label{eq:DifOpCreaAniqVac}
\begin{split}
\sqrt{2E_{p}} \left[ \hat{a}_{p, \lambda} (\infty) - \hat{a}_{p, \lambda}(-\infty) \right] & = -i \int d^{4}x \; e^{i p \cdot x} \varepsilon^{(-) *}_{\mu, \lambda}(p) \left[ \partial_{t}^{2} + E_{p}^{2} \right] \hat{W}^{(-) \mu}(x), 
\end{split}
\end{equation}
and, similarly,
\begin{equation}
\begin{split}
\sqrt{2E_{p}} \left[ \hat{a}^{\dagger}_{p, \lambda} (\infty) - \hat{a}^{\dagger}_{p, \lambda}(-\infty) \right] & = i \int d^{4}x \; e^{-i p \cdot x} \varepsilon_{\mu, \lambda}^{(-)}(p) \left[ \partial_{t}^{2} + E_{p}^{2} \right] \hat{W}^{(+) \mu}(x),
\end{split}
\end{equation}
the differences of asymptotic operators in a form suitable for LSZ reduction formula.

At this point, the appearance of $\partial_{t}^{2}+E_{p}^{2}$ is natural from the projection onto positive frequency plane waves. The next step is to relate this operator to the covariant equation of motion. Thus,  after performing the Fourier analysis of the Proca Equation and multiplying by $\varepsilon^{(-) *}_{\mu, \lambda}(p)$, we get
\begin{equation}
\int d^{4}x \; e^{i p \cdot x} \varepsilon_{\mu, \lambda}^{(-) *}(p) \left( \partial^{\alpha} \partial_{\alpha} + M^{2} \right) \hat{W}^{(-) \mu}(x) = \int d^{4}x \; e^{i p \cdot x} \varepsilon_{\mu, \lambda}^{(-) *}(p) \left( \partial_{t}^{2} + E_p^{2} \right) \hat{W}^{(-) \mu}(x),
\end{equation}
where $\partial^{\alpha}\partial_{\alpha}=\partial_{t}^{2}-\nabla^{2}$ was used, and two integrations by parts over the space were performed. Combining this identity with Eq.~\eqref{eq:DifOpCreaAniqVac} yields
\begin{equation}\label{aaainf}
\int d^{4}x \; e^{i p \cdot x} \varepsilon^{(-) *}_{\mu, \lambda} (p) \left( \partial^{\alpha} \partial_{\alpha} + M^{2} \right) \hat{W}^{(-) \mu}(x) = i \sqrt{2 E_{p}} \left[ \hat{a}_{p, \lambda}(\infty) - \hat{a}_{p, \lambda}(- \infty) \right],
\end{equation}
and similarly for the creation operator.

Substituting Eq.~\eqref{aaainf} together with its adjoint into Eq.~\eqref{eq:AmpProb}, and identifying the time-ordered two-point function, we obtain the LSZ reduction formula in the vacuum, given by
\begin{equation} \label{eq:LSZVacio}
\begin{split}
\bra{f} S \ket{i} & = \int d^{4}x \int d^{4}y \; e^{i p_{f} \cdot x} e^{-i p_{i} \cdot y} \varepsilon^{(-) *}_{\mu, \lambda_{f}}(p_{f}) \varepsilon^{(-)}_{\nu, \lambda_{i}}(p_{i}) \left(p_{f}^{2} - M^{2} \right) \left( p_{i}^{2} - M^{2} \right) \bra{\Omega} \mathcal{T} \left\{ \hat{W}^{(-) \mu}(x) \hat{W}^{(+) \nu}(y) \right\} \ket{\Omega}.
\end{split}
\end{equation}
where we have used that, upon Fourier transformation, the differential operators acting on the external points become the usual amputation factors $(p^{2}-M^{2})$.

The Eq.~\eqref{eq:LSZVacio} expresses the transition amplitude in terms of the time-ordered Green's function with the external propagators removed. In practical perturbative computations, the correlator
$\bra{\Omega}\mathcal{T}\{\hat{W}^{(-)\mu}(x)\hat{W}^{(+)\nu}(y)\}\ket{\Omega}$
generates Feynman diagrams containing internal propagators with denominators of the form $(p^{2}-M^{2}+i\varepsilon)^{-1}$. The factors $(p_{f}^{2}-M^{2})$ and $(p_{i}^{2}-M^{2})$ in Eq.~\eqref{eq:LSZVacio} precisely cancel these external denominators, thereby implementing the amputation of external legs and leaving the properly normalized on-shell amplitude. This vacuum derivation serves as the direct analog of the magnetic-field case developed in Sec.~\ref{sec.LSZMag}, where plane waves are replaced by Ritus eigenfunctions and $(p^{2}-M^{2})$ is replaced by the appropriate eigenvalue of the momentum operator in the presence of the background field.

\section{``Non-Rotated" Ritus Eigenfunction Representation} \label{sec.RitusDesRot}

As discussed by the end of Sec.~\ref{sec.WMF}, it is often convenient to express the charged vector boson propagator in a ``non-rotated'' basis, i.e., in the original basis where the spin structure is not diagonalized by the matrix $N_{\alpha}{}^{\mu}$ in Eq.~\eqref{eq:RotMat}. This representation is particularly useful in two contexts. First, it makes  the comparison with results commonly reported in the literature easier, where the diagonal basis is not always adopted. Second, it proves advantageous in Sec.~\ref{sec.LSZMag}, where we explicitly construct the LSZ reduction formula in a magnetic background, and it is natural to work with ``non-rotated'' external polarizations and fields.

To go from rotated to ``non-rotated'' eigenfunctions, let us start by recalling that the rotated field and the corresponding rotated polarization vectors are defined by $\tilde{W}^{(-)\mu}\equiv N^{\mu}{}_{\alpha}W^{(-)\alpha}$ and  $\tilde{\bar{\varepsilon}}_{\lambda}^{(-)\mu}\equiv N^{\mu}{}_{\alpha}\bar{\varepsilon}_{\lambda}^{(-)\alpha}$, respectively. Thus, it is not hard to see that the corresponding `non-rotated'' eigenfunctions can be obtained as
\begin{equation} \label{eq:EigenFunDesRot}
\begin{split}
\left( \mathbb{E}^{+}_{\ell , c} \right)^{\! \! \mu}_{\, \alpha} \!  (p, x) & \equiv
\left(N^{-1}\right)^{\mu}{}_{\nu}\,
\left( \tilde{\mathbb{E}}^{+}_{\ell , c} \right)^{\! \! \nu}_{\, \beta} \! \!  (p, x) \,  N^{\beta}{}_{\alpha}
\\
& \equiv \sum_{s_{3}} X^{+}_{\ell - 1 + c \hat{s}_{3}} (\rho_{+}) \Delta^{\mu}{}_{\alpha}(s_{3}) e^{-i \left( p^{0} x_{0} + p^{2} x_{2} + p^{3} x_{3} \right)}, \\
\end{split}
\end{equation}
where 
\begin{equation}
\begin{split}
 \Delta^{\mu}{}_{\alpha}(s_{3}) &\equiv \left( N^{-1} \right)^{\mu}{}_{\sigma} \tilde{\Delta}^{\sigma}{}_{\alpha}(s_{3}) N^{\alpha}{}_{\delta} \\
 &= \begin{pmatrix}
    1 - s_{3}^{2} & 0 & 0 & 0 \\
    0 & \frac{s_{3}^{2}}{2} & \frac{-i s_{3}}{2} & 0  \\
    0 & \frac{i s_{3}}{2} & \frac{s_{3}^{2}}{2} & 0 \\
    0 & 0 & 0 & 1 - s_{3}^{2}
\end{pmatrix},
\end{split}
\end{equation}
with $\tilde{\Delta}^{\mu}{}_{\nu}(s_{3})$ the diagonal spin projector in Eq.~\eqref{eq:MatProy}. Note that $\Delta^{\mu}{}_{\alpha}(s_{3})$ is no longer diagonal for $s_{3}=\pm1$, reflecting the fact that in the original basis the magnetic field couples the transverse components through the spin operator. Nevertheless, the projectors satisfy the same completeness relation,
\begin{equation}
\sum_{s_{3}=\pm1,0}\Delta^{\mu}{}_{\nu}(s_{3})=g^{\mu}{}_{\nu},
\end{equation}
and therefore, it is a consistent decomposition.

\noindent The above ideas can be easily extended to the quantized field operators of Eq. (\ref{eq:CuantizacionMag}); thus, the ``non-rotated'' quantized field operators are given by
\begin{equation} \label{eq:FourierDesRot}
\begin{split}
\hat{W}^{(-) \mu}(x) & = \SumInt \frac{d^{3}p}{(2 \pi)^{3}} \frac{1}{\sqrt{2E_{\bar{p}}}} \sum_{\lambda = 1}^{3} \left[ \hat{a}_{\bar{p}, \lambda} \left( \mathbb{E}^{+}_{\ell , 1} \right)^{\! \! \mu}_{\, \nu} \! (p, x) \bar{\varepsilon}_{\mathbf{\lambda}}^{(-) \nu}\left( \bar{p} \right) + \hat{b}_{\bar{p}, \lambda}^{\dagger} \left( \mathbb{E}^{-}_{\ell , 1} \right)^{\! \! \mu}_{\, \nu} \! (p, x) \bar{\varepsilon}_{\mathbf{\lambda}}^{(+) \nu *}\left( \bar{p} \right) \right], \\
\hat{W}^{(+) \mu}(x) & = \SumInt \frac{d^{3}p}{(2 \pi)^{3}} \frac{1}{\sqrt{2E_{\bar{p}}}} \sum_{\lambda = 1}^{3} \left[ \hat{a}^{\dagger}_{\bar{p}, \lambda} \left( \mathbb{E}^{+ \dagger}_{\ell, 1} \right)^{\! \! \mu}_{\, \nu} \! (p, x) \bar{\varepsilon}_{\mathbf{\lambda}}^{(-) \nu *}\left( \bar{p} \right) + \hat{b}_{\bar{p}, \lambda} \left( \mathbb{E}^{- \dagger}_{\ell, 1} \right)^{\! \! \mu}_{\, \nu} (p,x) \bar{\varepsilon}_{\mathbf{\lambda}}^{(+) \nu}\left( \bar{p} \right)  \right].
\end{split}
\end{equation}

The Feynman propagator follows from Eq.~\eqref{eq:FourierDesRot} through standard time-ordering and the application of the commutation relations. In practice, it is algebraically simpler to obtain it by rotating back the diagonal expression in Eq.~\eqref{eq:PropagadorMagnetico}, as follows
\begin{equation} \label{eq:PropagDesrotRitus}
\begin{split}
D^{\mu \nu}_{B}(x,y) & = (N^{-1})^{\mu}{}_{\sigma}  \tilde{D}_{B}^{\sigma \rho}(x,y) N_{\rho}{}^{\nu} 
\end{split}
\end{equation}
so, we arrive at
\begin{equation}
\begin{split}
D^{\mu \nu}_{B}(x,y) &=  i \SumInt \frac{d^{4}p}{(2 \pi)^{4}} \left( N^{-1} \right)^{\mu}{}_{\sigma} \frac{ \left( \tilde{\mathbb{E}}^{+}_{\ell , 1} \right)^{\! \! \sigma}_{\, \alpha} \! (p, x) \left[ - g^{\alpha \beta} + \left( 1 - \delta_{\ell, 0} \right) \frac{\tilde{\bar{p}}^{\alpha} \tilde{\bar{p}}^{\beta *}}{m_{W}^{2}} \right] \left( \tilde{\mathbb{E}}^{+ *}_{\ell , 1 } \right)_{\! \! \beta}^{\, \rho} \! (p, y)}{(p^{0})^{2} - E^{2} + i \varepsilon} N_{\rho}{}^{\nu} \\
& = i \SumInt \frac{d^{4}p}{(2 \pi)^{4}} \frac{ \left( \mathbb{E}^{+}_{\ell , 1} \right)^{\! \! \mu}_{\, \alpha} \! (p, x) \left[ - g^{\alpha \beta} + \left( 1 - \delta_{\ell, 0} \right) \frac{\bar{p}^{\alpha} \bar{p}^{\beta *}}{m_{W}^{2}} \right] \left( \mathbb{E}^{+ \dagger}_{\ell, 1} \right)_{\! \! \beta}^{\, \nu} \! (p, y) }{(p^{0})^{2} - E^{2} + i \varepsilon},
\end{split}
\end{equation}
which is the ``non-rotated" Ritus eigenfunction representation of the propagator. Here $\bar{p}^{\mu}$ denotes the ``non-rotated'' magnetic four-momentum obtained by transforming $\tilde{\bar{p}}^{\mu}$ back to the Lorentz basis, 
\begin{equation}
\begin{split}
\bar{p}^{\mu} 
&\equiv
\left(N^{-1}\right)^{\mu}{}_{\alpha}\,\tilde{\bar{p}}^{\alpha}\\
& = \Bigg( p^{0}, - i s_{eB} \sqrt{\frac{|eB|}{2}} \left[ \sqrt{\ell_{+}} - \sqrt{ \ell_{-}} \right], s_{eB} \sqrt{\frac{|eB|}{2}} \left[ \sqrt{ \ell_{+}} + \sqrt{ \ell_{-}} \right], p^{3} \Bigg).
\end{split}
\end{equation}
The representation \eqref{eq:PropagDesrotRitus} is structurally similar to the vacuum propagator: the denominator shows a pole structure, while the background magnetic field dependence is encoded in the eigenfunctions $\mathbb{E}_{\ell}$ and in the momentum $\bar{p}^{\mu}$. This is precisely the feature that makes the Ritus approach well suited for LSZ-type amputations, as developed in Sec.~\ref{sec.LSZMag}. We also note that a closely related expression to Eq.~\eqref{eq:PropagDesrotRitus} appears in Ref.~\cite{ChargedMeson_scoccola}; however, the definitions of the spin projectors and the momentum $\bar{p}^{\mu}$ differ, reflecting alternative conventions for the ``non-rotated'' basis and for the implementation of the spin decomposition in the presence of the magnetic field.

\section{On the $\hat{\tilde{\Pi}}_{+}$ acting in Eq.~\eqref{eq:PreSubsecscovv}} \label{piapendix}

In order to make analytical progress, let us introduce the notation
\begin{equation}
\begin{split}
& \mathscr{E} \equiv\Omega(x,y) e^{-i  p\cdot\left( x - y \right)_{\parallel}} e^{\frac{ieB}{4} \cot{(eBs)} \left[ \left( x - y \right)_{\mu} \varphi^{\mu \sigma} \varphi_{\sigma}{}^{\nu} \left( x - y \right)_{\nu} \right]},\\    
\end{split}
\end{equation}
as the whole exponential factor over which the operator $\hat{\tilde{\Pi}}^{\beta *}_{+}(y)$ is applied. 

The result, once the operator has been applied, is 
\begin{equation}
\begin{split}
\hat{\tilde{\Pi}}^{\beta *}_{+}(y) \mathscr{E} = -i \mathscr{E} \left[ i p^{\beta}_{\parallel} - i \frac{eB}{2} \left( \mathbb{B}^{\beta} (y,x) \right)^{*} - i \frac{eB}{2} \cot(eBs) \left( \mathbb{C}^{\beta} (x,y) \right)^{*} + \tilde{\mathscr{A}}^{\beta}_{y} \right],
\end{split}
\end{equation}
where
\begin{equation}
\begin{split}
&\tilde{\mathscr{A}}^{\alpha}_{x} \equiv -ie A^{\mu}(x) \left( N^{-1} \right) _{\mu}{}^{\alpha}, \\
& \mathbb{B}^{\alpha}(x,y) \equiv N^{\alpha}{}_{\mu} \varphi^{\mu \nu} y_{\nu} + \frac{i}{2} \tilde{\varphi}^{\alpha}{}_{\mu} \left[ \left( N^{\mu}{}_{\nu} \right)^{*} - N^{\mu}{}_{\nu} \right] \varphi^{\nu \rho} \varphi_{\rho}{}^{\sigma} x_{\sigma} + N^{\alpha}{}_{\mu} \varphi^{\mu}{}_{\nu} \varphi^{\nu \rho} \frac{A_{\rho}(x)}{B}, \\
& \mathbb{C}^{\alpha}(x,y) \equiv N^{\alpha}{}_{\mu} \varphi^{\mu}{}_{\nu} \varphi^{\nu \rho} \left( x - y \right)_{\rho}.
\end{split}
\end{equation}

Now, by applying both differential operators to the exponential factors, we get
\begin{equation}
\begin{split}
&\hat{\tilde{\Pi}}^{\alpha}_{+}(x) \hat{\tilde{\Pi}}^{\beta *}_{+}(y)\mathscr{E}  \\
&\hspace{2em} = \mathscr{E} \Bigg\{ \Bigg[ -i p^{\alpha}_{\parallel} + i \frac{eB}{2} \left[  \mathbb{B}^{\alpha}(x,y) + \cot(eBs) \mathbb{C}^{\alpha}(x,y) \right] + \tilde{\mathscr{A}}^{\alpha}_{x} \Bigg] \Bigg[ i p^{\beta}_{\parallel} - i \frac{eB}{2} \left[  \mathbb{B}^{\beta}(y,x) + \cot(eBs) \mathbb{C}^{\beta}(x,y) \right]^{*} + \tilde{\mathscr{A}}^{\beta}_{y} \Bigg] \\
&\hspace{6em} + \frac{ieB}{2} \left[ N^{\alpha}{}_{\mu} \varphi^{\mu \nu} \left( N^{-1} \right)_{\nu}{}^{\beta} - \cot(eBs) N^{\alpha}{}_{\mu} \varphi^{\mu \nu} \varphi_{\nu}{}^{\rho} \left( N^{-1} \right)_{\rho}{}^{\beta} \right] \Bigg\}.
\end{split}
\end{equation}

Additionally, in deriving Eq.~\eqref{eq:InvariantProp1} we made use of the identities
\begin{equation}
\begin{split}
& e^{i \frac{eB}{4} \cot{(eBs)} (X_{i})^{2}} = \int^{\infty}_{-\infty} dp^{i} \; e^{-i p^{i} X_{i}} e^{-i (p^{i})^{2} \frac{\tan{(eBs)}}{eB}} \sqrt{\frac{i \tan{(eBs)}}{\pi eB}}, \\
& X_{i} e^{i \frac{eB}{4} \cot{(eBs)} (X_{i})^{2}} = \int^{\infty}_{-\infty} dp^{i} \; e^{-i p^{i} X_{i}} e^{-i (p^{i})^{2} \frac{\tan{(eBs)}}{eB}} \frac{2i}{\sqrt{\pi}} \left( \frac{i \tan{(eBs)}}{eB} \right)^{\frac{3}{2}}, \\
& (X_{i})^{2} e^{i \frac{eB}{4} \cot{(eBs)} (X_{i})^{2}} = \int^{\infty}_{-\infty} dp^{i} \; e^{-i p^{i} X_{i}} e^{-i (p^{i})^{2} \frac{\tan{(eBs)}}{eB}} \frac{-2i}{eB \cot{(eBs)}} \Bigg[ (p^{i})^{2} \frac{2}{\sqrt{\pi}} \left( \frac{i \tan{(eBs)}}{eB} \right)^{\frac{3}{2}} - \sqrt{\frac{i \tan{(eBs)}}{\pi eB}} \Bigg],
\end{split}
\end{equation}
which allow us to rewrite Gaussian factors in coordinate space as Gaussian integrals in momentum space.
In the above expressions $X_{i} \equiv x_{i} - y_{i}$.

\bibliographystyle{unsrt}
\bibliography{Articulo_WBosonMagneticField}

\end{document}